\documentclass[twocolumn]{IEEEtran}

\usepackage{times,amsmath,epsfig,graphpap,graphics,cite,amsfonts,amssymb}
\usepackage{makecell, multirow, url,hyperref,color}

\def\BibTeX{{\rm B\kern-.05em{\sc i\kern-.025em b}\kern-.08em
    T\kern-.1667em\lower.7ex\hbox{E}\kern-.125emX}}

\usepackage{epsfig,times,amsfonts,graphicx,color,amssymb,setspace}

\title{ITEM: Immersive Telepresence for Entertainment and Meetings - A Practical Approach} 
\author{Viet-Anh Nguyen, Jiangbo Lu, {\em Member, IEEE}, Shengkui Zhao, Tien Dung Vu, Hongsheng Yang, \\ Jones L. Douglas, {\em Fellow, IEEE}, and Minh N. Do, {\em Fellow, IEEE} \thanks{This study is supported by the research grant for the Human Sixth Sense Programme at the Advanced Digital Sciences Center from Singapore's Agency for Science, Technology and Research (A*STAR).}
}

\begin{document}
\maketitle
\begin{abstract}
This paper presents an Immersive Telepresence system for Entertainment and Meetings (ITEM). The system aims to provide a radically new video communication experience by seamlessly merging participants into the same virtual space to allow a natural interaction among them and shared collaborative contents. With the goal to make a scalable, flexible system for various business solutions as well as easily accessible by massive consumers, we address the challenges in the whole pipeline of media processing, communication, and displaying in our design and realization of such a system. Particularly, in this paper we focus on the system aspects that maximize the end-user experience, optimize the system and network resources, and enable various teleimmersive application scenarios. In addition, we also present a few key technologies, i.e. fast object-based video coding for real world data 
and spatialized audio capture and 3D sound localization for group teleconferencing. Our effort is to investigate and optimize the key system components and provide an efficient end-to-end optimization and integration by considering user needs and preferences. Extensive experiments show the developed system runs reliably and comfortably in real time with a minimal setup requirement (e.g. a webcam and/or a depth camera, an optional  microphone array, a laptop/desktop connected to the public Internet) for teleimmersive communication. With such a really minimal deployment requirement, we present a variety of interesting applications and user experiences created by ITEM.
\end{abstract}


\sloppy
\section{Introduction}
\label{sec:intro}
 Face-to-face meeting has been necessary for effective communication, but time, financial cost, and also environmental concerns are becoming less tolerable. With the advent of network and multimedia technologies, virtual meeting has become increasingly popular to enable more frequent and less costly person-to-person communication. However, most traditional virtual meeting systems in both business and consumer spaces such as Skype or Cisco WebEx still provide rather limited and sometimes unsatisfying functionalities and hardly maintain the experience of an in-person meeting. For instance, the separate displays of remote participants fail to provide a sense of co-location as in face-to-face meetings, while their poor integration with the shared collaborative contents and virtual environments leads to limited non-verbal cues and interaction among them.

To address these issues, tele-immersive (TI) systems with the vision of providing natural user experiences and interaction have attracted many research interests in the last decade~\cite{TIreview}. High-end telepresence products such as Cisco TelePresence~\cite{cisco} and HP's Hallo are designed to create the perception of meeting in the same physical space, which demand a proprietary installation and high-cost setup. Recently, some 3D tele-immersive (TI) systems have been developed to enhance the remote collaboration by presenting remote participants into the same 3D virtual space~\cite{mm04,teevee-mm09,teeve05}. However, these systems still fall short of stimulating a face-to-face collaboration with the presence of shared contents. Also, requiring bulky and expensive hardware with nontrivial calibration and setup hinders their wide deployment in real life.
   \begin{figure} [pt] 
   \begin{center}
   \includegraphics[width=\columnwidth]{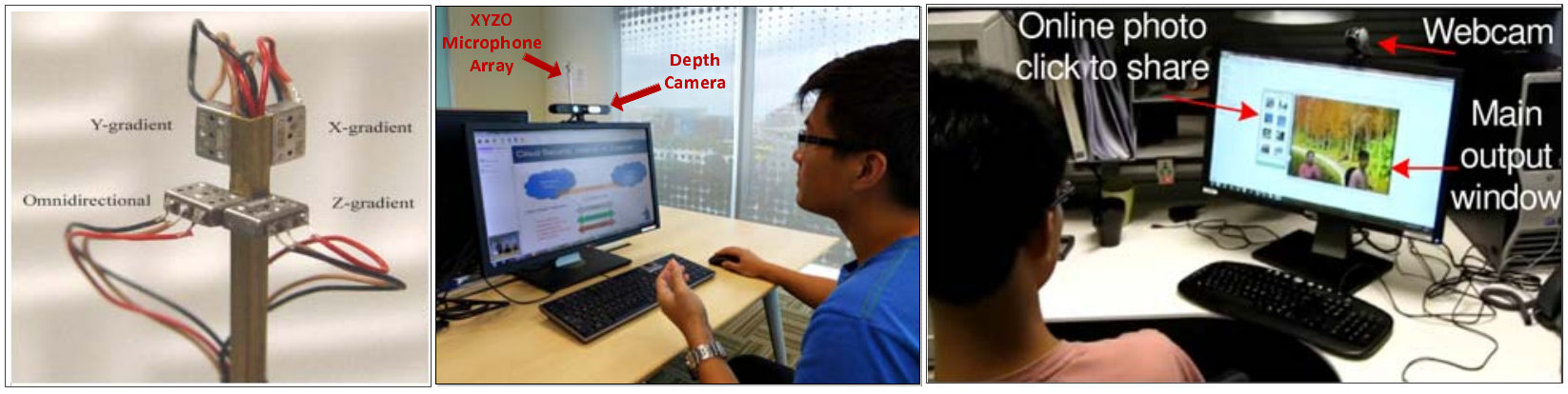}
   \end{center}
   \vspace{-8pt}
    \caption
   {\label{fig:fig1}
   Required hardware and various setups of the ITEM system. (a)~Microphone array. (b)~ITEM system setup with microphone array and depth camera. (c)~ITEM system setup with webcam. (All the figures in this paper are best viewed in color.)}
   \vspace{-18pt}
   \end{figure}

Motivated by these challenges and more, we present in this paper an Immersive Telepresence system for Entertainment and Meetings (ITEM) based on a low-cost, flexible setup (e.g. a webcam and/or a depth camera, an optional low-cost microphone array, a desktop/laptop connected to the public Internet). As shown in Fig.~\ref{fig:fig1}, the system allows putting two or more remote participants into the same virtual space and seamlessly integrating them with any shared collaborative contents for a more natural person-to-person interaction. With an addition of a depth camera and a low-cost microphone array, ITEM supports spatialized 3D audio, active speaker detection, and gesture-based controls to reproduce nonverbal signals and interactions with the shared contents in an intuitive and effective manner.

This paper describes a complete design and implementation of such a system by addressing the challenges in the whole pipeline of media processing, communication, and displaying. We consider major practical requirements in our design to build a system that supports \textit{multimodality} (audio/video, shared contents), \textit{scalability} for a large number of participants and concurrent meetings, \textit{flexibility} in a system setup (2D color webcam and/or 3D depth camera), and desirable \textit{functionality} to best suit different application contexts. Taking a systematic and integrated approach, we seamlessly combine various key components from the high-quality video object cutout, efficient media coding delivery, 3D sound source capture and localization, and immersive composition to improve system performance and enhance immersive experience. As an end result, we present several interesting applications and user experiences created by ITEM. 

In summary, we particularly focus on the system aspects in building such a lightweight practical TI system that maximizes the end-user experience, optimizes the system and network resources, and enables a variety of teleimmersive application scenarios. Our major effort is to investigate and optimize the key system components, individually and collectively, and  efficient end-to-end integration by taking user needs and preferences into consideration. The main contributions of this work can be summarized as follows
\begin{itemize}
	\item An end-to-end optimized practical TI system using inexpensive and widely available hardware (e.g. common video peripherals and CPUs)
	\item A fine-tuned object-based fast video coding technique for real world data
	\item Spatialized audio capture and 3D sound localization using a low-cost microphone array for group teleconferencing
	\item Detailed objective and subjective evaluation for extensive user study of this first lightweight TI system. 
\end{itemize}

As a result, the proposed ITEM system advances the state of the art TI systems in four major and unique ways 
\begin{itemize}
	\item A low-cost and simple setup for minimal deployment requirement of a real-time TI system.
	\item Stimulate immersive experience in the same environment and sense of being together for remote participants.
	\item New user experiences in group teleconferencing created by the microphone array and its seamless integration with the visual processing.
	\item Reduce the computation load and bandwidth requirement by a factor of 3 to 4 compared to existing TI systems.
\end{itemize}
The remainder of the paper is organized as follows. Section II presents the literature review of the state-of-the-art immersive telepresence systems. 
We present in Section III the overview and architecture design of the proposed ITEM system, while the key developed technologies are discussed in Section IV. Section V presents the proposed fast object based coding component of the ITEM system. In Section VI, we present the proposed 3D audio capturing and localization technology for group teleconferencing. The system performances are shown in Section VII. In Section VIII, we briefly discuss several immersive applications and user experiences created by ITEM. Section IX provides the concluding remarks. A preliminary version of this work has been presented in~\cite{item,objcoding,cutechat}.

\section{Related Work}

\begin{table*} [t]
\centering
\small
\vspace{-4pt}
\caption{Comparison between ITEM and the existing video teleconferencing solutions.}
\label{tab:review}
\begin{tabular}{|l|c|l|c|l|l|} \hline
\multicolumn{1}{|c}{Solutions} &	\multicolumn{1}{|c}{Setup cost} &	\multicolumn{1}{|c}{Hardware} &	\multicolumn{1}{|c}{Network}	& \multicolumn{1}{|c}{Audio/video quality} & \multicolumn{1}{|c|}{QoE} \\ \hline
High-end Teleprecense  & Extremely & Dedicated setup,	& Dedicated 	& Life-size video quality, & Immersive illusion, \\ 
(Cisco~\cite{cisco}, HP's Halo) & high & propriety hardware	&bandwidth  & studio room quality & peripheral awareness \\ \hline

NTII~\cite{mm04},TEEVE~\cite{teeve05} & High & Bulky, expensive	& Internet2 & Reliable video cutout, & 3D immersive rendering,\\ 
BeingThere~\cite{Beingthere}& & 3D camera setup, & network & low 3D video quality, & body interaction and\\
& & blue screen &  & low frame rate, & collaboration\\ 
& &             &  & standard audio &			    \\ \hline

2D TI systems & Low & Standard PCs-audio, & Public   & Unreliable video cutout, & Immersive discussion,\\ 
(Virtual meeting~\cite{vms02}, & & video peripherals, & Internet  & low video resolution, & w/o supporting non-\\ 
Coliseum~\cite{coliseum}, & & multiple cameras  & & low frame rate, & verbal signals/cues \\
CuteChat~\cite{cutechat}) & & (in Coliseum)& &stand audio & and collaboration  \\ \hline

Standard video & Low & Standard PCs-audio, & Public  & Medium to high video & Non-immersive, w/o\\ 
conferencing (Skype) & &video peripherals &Internet & quality (HD resolution), & non-verbal collaboration \\ 
 & & & & standard audio & \\ \hline

ITEM & Low & Standard PCs-audio, & Public  & Robust, reliable cutout, & Immersive, natural\\ 
 & & video peripherals, &Internet & support HD resolution & conversation with\\
& & depth cam (optional), & & high frame rate, & non-verbal collaboration\\ 
& &microphone array  & & spatial audio,& \\ 
& &  (optional)  & & speaker detection& \\\hline
\end{tabular}
\vspace{-0.1in}
\end{table*}

Though standard-based video conferencing systems such as Skype or Cisco WebEx can provide a low-cost, easily accessible virtual meeting solution, they tend to impoverish the video communication experience due to not giving the feeling of being immerse and interactive. As a result, the ultimate goal of an immersive system is to enable natural experiences and interaction among remote participants from different locations to mimic in-person meetings as if they belong to the same physical space. Many efforts have been attempted to create such systems for achieving the two main targets: \textsl{immersive communication} for supporting natural conversation and \textsl{immersive collaboration} for sharing information.

To support immersive communication, high-end telepresence products such as Cisco TelePresence~\cite{cisco} and HP's Halo are sophisticatedly designed to provide participants with the perception of meeting in the same physical room through the ideal and life-size displays. Cisco system also simulates spatialized audio by rendering the sound of remote participants from specific locations rather than creating a truly 3D immersive audio. However, this is usually done with a number of permanently installed loudspeakers. These systems generally demand a proprietary installation and high-cost setup such as specific room design with advanced equipments and network infrastructures to maintain the illusion of a single room and support real-time, low-latency communication for an immense amount of media data. Several systems with spatialized audio~\cite{3DAudio} typically assume one participant at each station and do not offer 3D audio capture; therefore, they are ineffective in group teleconferencing. In order to have a better intelligibility and an easier comprehension in such scenarios, several attempts were explored by utilizing the sound source localization (SSL) and visual information to automatically detect and track the active speaker. They often require a multiple and large spatially separated microphone setup~\cite{SSL2,SSL3, SSL1} or a specific stand-alone microphone array device (e.g. Microsoft RoundTable~\cite{roundtable}). Meanwhile, in response to the need of immersive interaction, some 3D tele-immersive (TI) systems, such as the National Tele-Immersion Initiative~\cite{mm04} and TEEVE~\cite{teeve05}, have been recently developed to address the lack of remote collaboration by immersiving 3D representations of remote participants into the same 3D virtual space. As a result, these systems allow interaction of body movement such as dancing.  However, they are usually very bulky with expensive 3D video capturing setup, bandwidth-demanding, and computationally intensive, which seriously limit their wide applications in daily practice. Though Wu et al.~\cite{teevee-mm09} recently attempted to enhance the portability of the existing TI systems, their system still demands nontrivial setups.

With the goal to make the system easily assessable by massive consumers, prior research has also been conducted to develop 2D
TI systems (e.g. \cite{vms02,coliseum,cutechat}) based on commodity cameras and computers. These systems share some similarity with our approach where the foreground objects are segmented from undesired background to allow merging two or multiple participants into the same shared background environment. As a result, participants can experience collaboration capability through natural interaction with each other or with the selected background content (e.g. presentation slides, online photo albums), which does not exist in the traditional video conferencing systems such as Skype. However, the key foreground segmentation techniques adopted are often very rudimentary~\cite{vms02} or computationally intensive {~\cite{coliseum}} and provide a quality-compromised segmentation performance. 

Though recent years have witnessed some important advances in the research field of foreground segmentation from a live video ~\cite{bi-cvpr06,bi-cvpr05,bgdcut06,tofcut10}, the technology existing today is still distant from what a practical solution really desires, especially when it is put under a context of live video teleconferencing. Some known issues are small video resolutions e.g. $320\times240$, inaccurate foreground segmentation under challenging test conditions, and requiring stereo cameras~\cite{bi-cvpr05} or additional depth sensors~\cite{tofcut10}.  Recently, Lu~{\em{et al.}}~\cite{cutechat} have developed a more advanced segmentation technique to realize a practical 2D TI system. Without supporting a flexible setup of incorporating a depth camera when available, it does not handle challenging situations well by using only a single webcam (e.g. background/foreground with very similar colors or difficult hand gestures). Furthermore, their system does not support multimodality and allows only limited interactions with shared contents, while the current design lacks capabilities to support many concurrent meeting sessions and flexibility in functionality extension in different application contexts.

In summary, Table~\ref{tab:review} lists the pros and cons in various areas of the existing video teleconferencing solutions and highlights the key and unique selling points of  our proposed system.


\section{ITEM SYSTEM}
\label{sec:sys}

\subsection{Overview}

   \begin{figure} [pt]
   \begin{center}
   \includegraphics[width = \columnwidth]{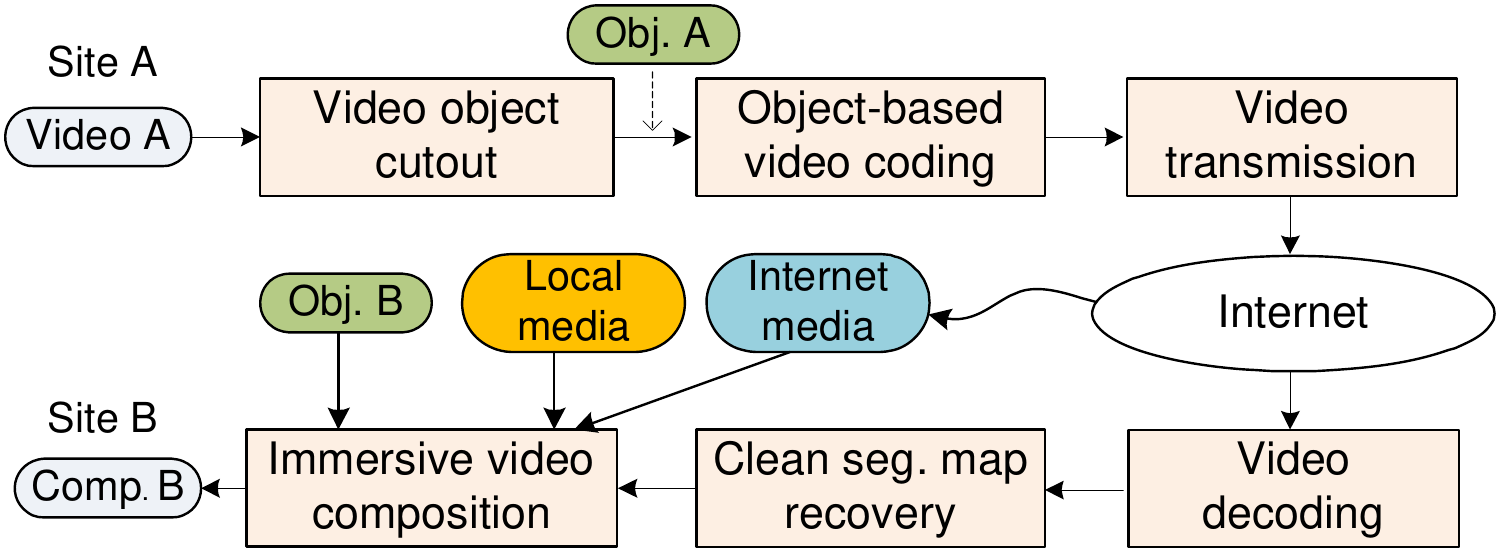}
   \end{center}
   \vspace{-8pt}
   \caption
   {\label{fig:fig2}
   System overview.}
    \vspace{-0.22in}
     \end{figure}

Fig.~\ref{fig:fig2} gives an overview of the ITEM system, where only a pair of a sender and a receiver is shown for simplicity. In reality, our system actually supports multiparty, two-way multimodal conferencing. At the sender site (site A in Fig.~\ref{fig:fig2}), a commodity webcam (or a depth camera) is used to capture a live video stream, which is processed with our video object cutout technique to segment out object A in real time. Next, object-based video coding is performed to encode the foreground object stream using blue color as the background chroma key. Then, the compressed object video is transmitted over the public Internet, reaching the receiver site under the management of an enhanced video delivery scheme. After decoding the object video, a clean segmentation map recovery method is applied to reconstruct a clean foreground segmentation map, which would otherwise contain boundary artifacts caused by compressing object video with background chroma keying. Lastly, the user has a range of options regarding how to composite the final video to be displayed on screen. She can choose to merge her own object video into the final frames, while the background content (e.g. slides, photos) can be selected either from the local store, or streamed dynamically from somewhere in the Internet as shared by the site A. 
Furthermore, in a group teleconferencing scenario, a low-cost compact microphone array is used to provide 3D audio capture and reproduction as well as active speaker detection and tracking. The new communication experiences and compelling functionalities created by ITEM can be seen, in the accompanying video, http://youtu.be/cuRvXcLUIR4.

\vspace{-4pt}
\subsection{System Architecture}
\begin{figure} [t]
   \begin{center}
    \includegraphics[width = \columnwidth]{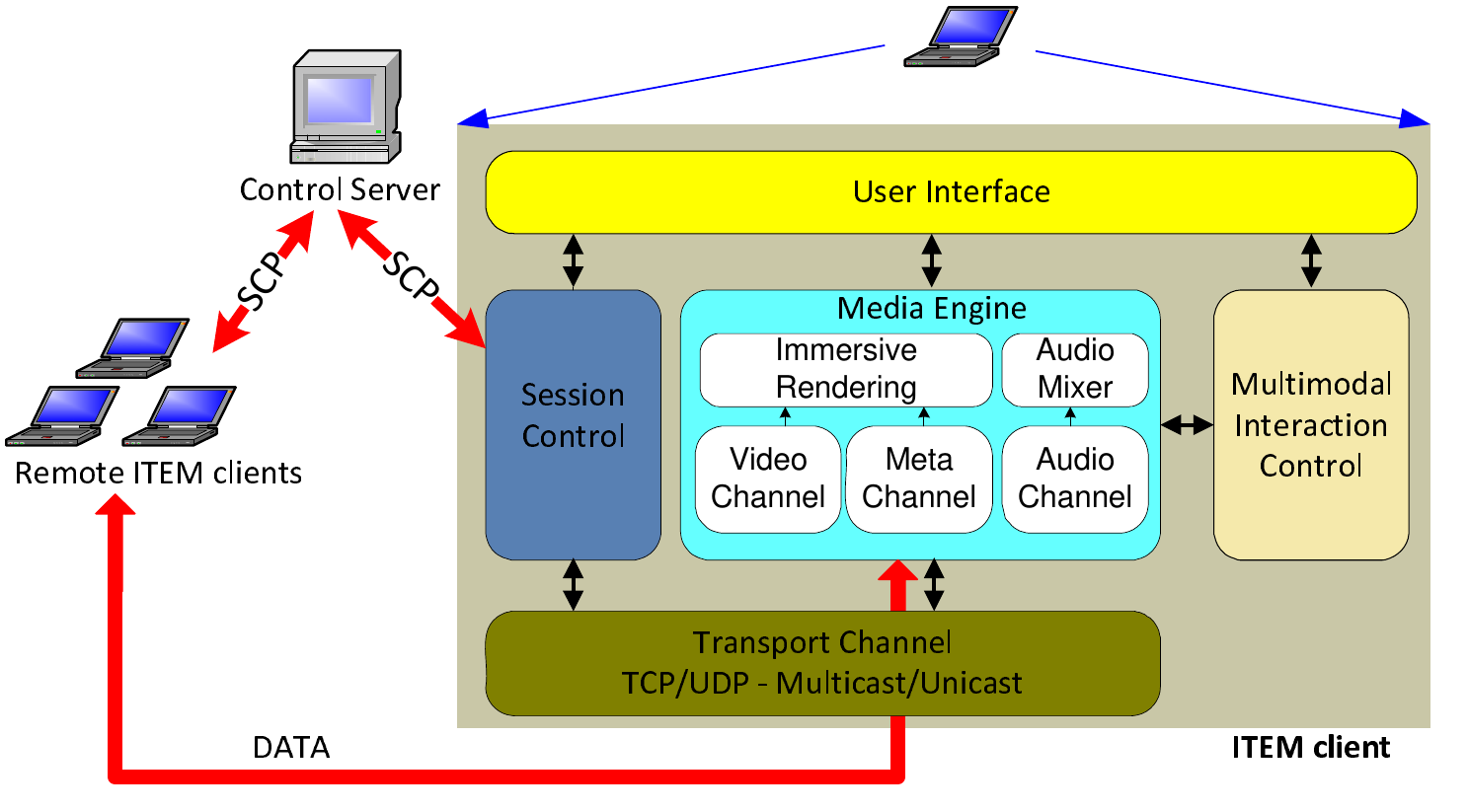}
   \end{center}
   \vspace{-10pt}
   \caption
    { \label{fig:fig3}   
      Simplified diagram of the system design.}
      \vspace{-8pt}
   \end{figure}

ITEM is designed not only to support basic requirements of a typical conferencing solution (e.g. high video/audio quality, ease of use), but also to provide compelling features for immersive experience. A modular design approach is employed in the realization of  this system to improve reusability, extensibility, and reconfigurability in various application contexts. Fig.~\ref{fig:fig3} depicts the simplified data flows and major components of an ITEM client.

   \begin{figure} [pt]
   \begin{center}
   \includegraphics[width = \columnwidth]{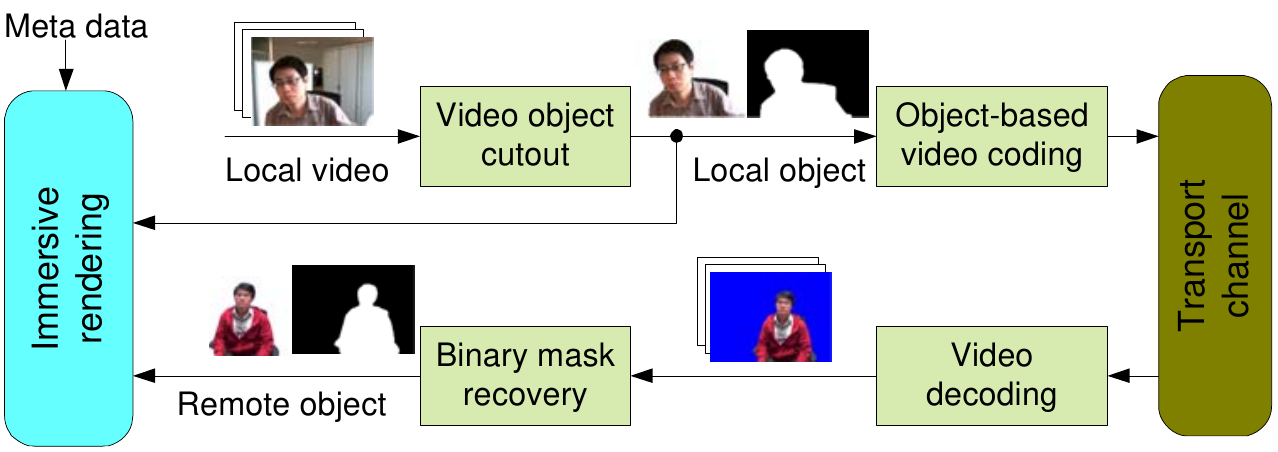}
   \end{center}
   \vspace{-10pt}
   \caption
   {\label{fig:vchannel}  
	 Block diagram and data flows of the video channel.}
	 \vspace{-15pt}
   \end{figure}

\subsubsection{Session control}
The session control module manages the initialization and control of a communication session including both resource information (e.g. media capabilities, transmission paths) and process management (e.g. initiation, termination) by communicating with a control server through session control protocol (SCP). Without handling media data, the light-weight control server easily supports many concurrent sessions. We design a simplified set of SCP to facilitate a client interaction with a session such as \textit{creation}, \textit{termination}, \textit{joining}, and \textit{leaving}. Our design provides the reconfigurability for various application scenarios by creating and storing in a control server the media transmission structure, channel properties (e.g. unicast or multicast), and participants' media capabilities for each session. An appropriate connection is established for a client that opts to join a session either as a {\it passive participant} by only receiving media data or as an {\it active participant} by also sharing her media data. The control server is also used for new user registration and login authentication through a membership database.   

\subsubsection{Media engine}
The role of media engine is to process both the local media prior to transmission and the incoming media from remote users for immersive composition. The engine provides seamless audio/video communication among users through a video/audio channel, while shared contents (e.g. documents, media-rich information) are processed through a meta channel. Regardless of channel types, video-audio contents are transmitted using RTP/RTCP protocol in order to support QoS (e.g. packet loss, jitter, latency, bandwidth adaptation) and audio/video synchronization. Basically, the meta channel inherits many similar modules from the audio/video channel to process shared audio/video contents, while other data types can also be easily handled through a secure protocol like TCP. To meet the low-latency requirement and to improve the system performance, multithreading techniques are employed to independently handle different channels and incoming data from each remote user. 

\textbf{Audio channel:} The audio channel accommodates 3D sound processing and SSL for spatialized audio and speaker detection to enhance the user experiences in group teleconferencing. Audio stream is processed in this channel to provide audio capability that is not supported in~\cite{cutechat}. We use G.711 codec for audio compression, while a wide range of audio codecs are supported (e.g. G.729, Speex, etc.) for decoding the incoming audio streams from shared contents. A simple audio mixer is also provided to merge audio streams from multiple sources and synchronize with video streams using information from RTCP packets.

\textbf{Video channel:} This is the most critical module in ITEM that enables a variety of functional features for desired immersive experience by processing the user video on an object basis. Fig.~\ref{fig:vchannel} shows the simplified processing flow of the video channel. Our \textit{\textbf{video object cutout}} technology is first applied to segment the user object in real time from a live video captured by a single webcam or a depth camera such as Microsoft Kinect. For efficient delivery, a novel and fast \textit{\textbf{object-based video coding}} is proposed to encode the foreground object stream using a chroma-key-based scheme with H.264 codec, which will be discused shortly.

\begin{figure} [t]
  
   \begin{center}
   \includegraphics[width = \columnwidth]{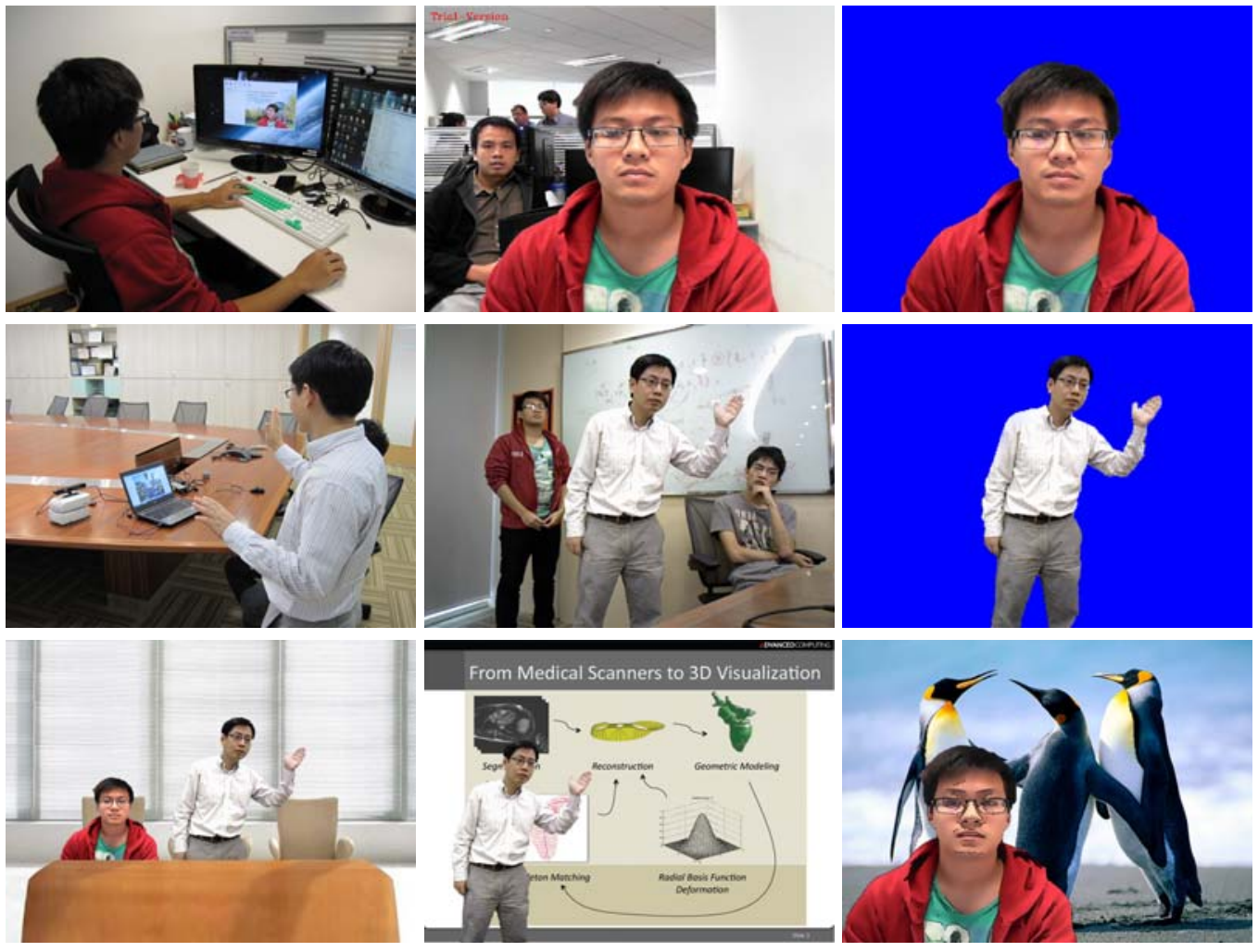}
   \end{center}
     \vspace{-6pt}
   \caption
   {\label{fig:cutout}  
	 ITEM's real-time video object cutout technology using a normal webcam (top row) or a depth (plus RGB) camera (middle row). From left to right: system setup, input video frame, and cutout result. Bottom row shows some video composition effects.
   }
    \vspace{-16pt}
   \end{figure}

\textbf{Immersive rendering:} This module is responsible for merging user objects from different sources with shared contents from the meta channel in an immersive and interactive manner (see Fig.~\ref{fig:cutout}). For low-latency renderization, it is desired to refresh the composed frame upon receiving new data from any sources. With multiple asynchronous video streams, such a renderization strategy may overload the CPU usage due to a high rendering frame rate incurred. Thus, we use a master clock to update and render the composed frame at some frame rate (e.g. 30 FPS) without introducing any noticeable delay. For the ease of object manipulation, an object index map is used to indicate the object location in the composed frame for each media source.    


\subsubsection{Multimodal interaction control}
For a more natural interaction with the shared contents, the cumbersome of using a keyboard and a mouse should be avoided whenever appropriate. With an addition of a depth camera, we employ hand gestures to interact with the system and provide users a comfortable lean back experience. Currently, we support several hand gestures to control the shared contents (e.g. paging through the slides). For more reliable tracking of hand motions, we have developed a fast hand gesture detection approach, which performs consistently better than the baseline OpenNI SDK, also in a more responsive manner.


\subsubsection{Transport channel}
Transport channel communicates with the session control and media engine modules to create an appropriate connection for data transmission in various channels based on the transmission architectures and data types. The module assigns and manages the list of destinations (e.g. a remote user address or a multicast address, if available). Real-time audio/video data is transmitted using UDP protocol in RTP/RTCP packetization. Meanwhile, SCP data and other types of shared data in the meta channel such as text are transmitted using TCP protocol for reliable transmission.  


\section{Key Technologies}
In what follows, we shall present the key developed technologies highlighted earlier in Section III, which make the realization of the ITEM system possible. 

\vspace{-8pt}
\subsection{Video Object Cutout}
\begin{figure} [t]
   \begin{center}
   \includegraphics[width = \columnwidth]{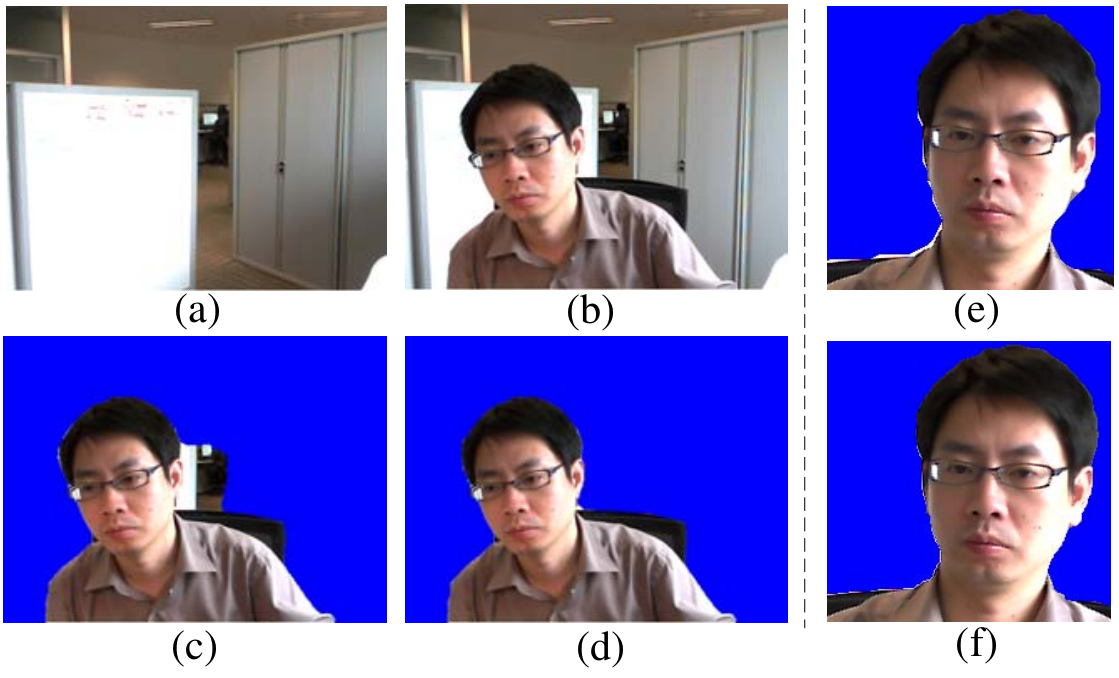}
   \end{center}
   \vspace{-12pt}
   \caption
   {\label{fig:cutout1}
   Video object cutout. (a) Known background. (b) A live video frame where a background person with similar colors was moving. (c) Cutout result yielded by our implementation of \cite{bgdcut06}. (d) Our cutout result. Close-up of the result without (e) and with (f) using our labeling refinement technique.}
      \vspace{-2pt}
   \end{figure}

As illustrated in Fig.~\ref{fig:fig2}, video object cutout is the very first processing step that is critical to the proposed ITEM system. Assuming that the background is known and the webcam is static, we have developed a practical solution for real-time segmenting a foreground layer from a live video captured by a single webcam. Though this assumption appears somewhat constrained, a good solution can be widely deployed and it enables aforementioned exciting applications with no additional cost. In fact, segmenting the foreground layer accurately from a complex scene where various changes can happen in the background is still rather challenging. 
Compared with the state-of-the-art segmentation techniques, our technology has shown a few important advantages that make ITEM competitive and practical: 1) reliable segmentation with high accuracy under challenging conditions, 2) real-time speed (18-25 FPS for VGA-resolution, 14-18 FPS for HD-resolution) on a commodity hardware such as a laptop/desktop, and 3) ease of use with little or no user intervention in the initialization phase. The technology has been registered as a trade secret recently and a detailed description is out of the scope of this paper. Basically, based on a unified optimization framework, our technology probabilistically fuses different cues together with spatial and temporal priors for accurate foreground layer segmentation. In particular, the proposed technology consists of two major steps, i.e. {\it layer estimation} and {\it labeling refinement} (see Fig.~\ref{fig:cutout1}). When a depth camera is available, the current framework can also be automatically configured to utilize the important depth information for more reliable inference, while leveraging several other key components also shared by the webcam-based object cutout flow. Fig.~\ref{fig:cutout} shows more foreground segmentation results using different setups (e.g. with/without using a depth camera) under challenging test conditions. 

\vspace{-8pt}
\subsection{Object-Based Video Coding}
In ITEM, we propose to compress and deliver only foreground objects while the shared background can be streamed dynamically from the cloud. Using the high downlink bandwidth to synchronize the shared background from the cloud allows reducing the traffic otherwise in the uplink. We will show later that throwing away the undesired background can reduce the uplink bandwidth three times or more and ease the network congestion and delay.

\begin{figure} [t]
   \begin{center}
   \includegraphics[width =\columnwidth]{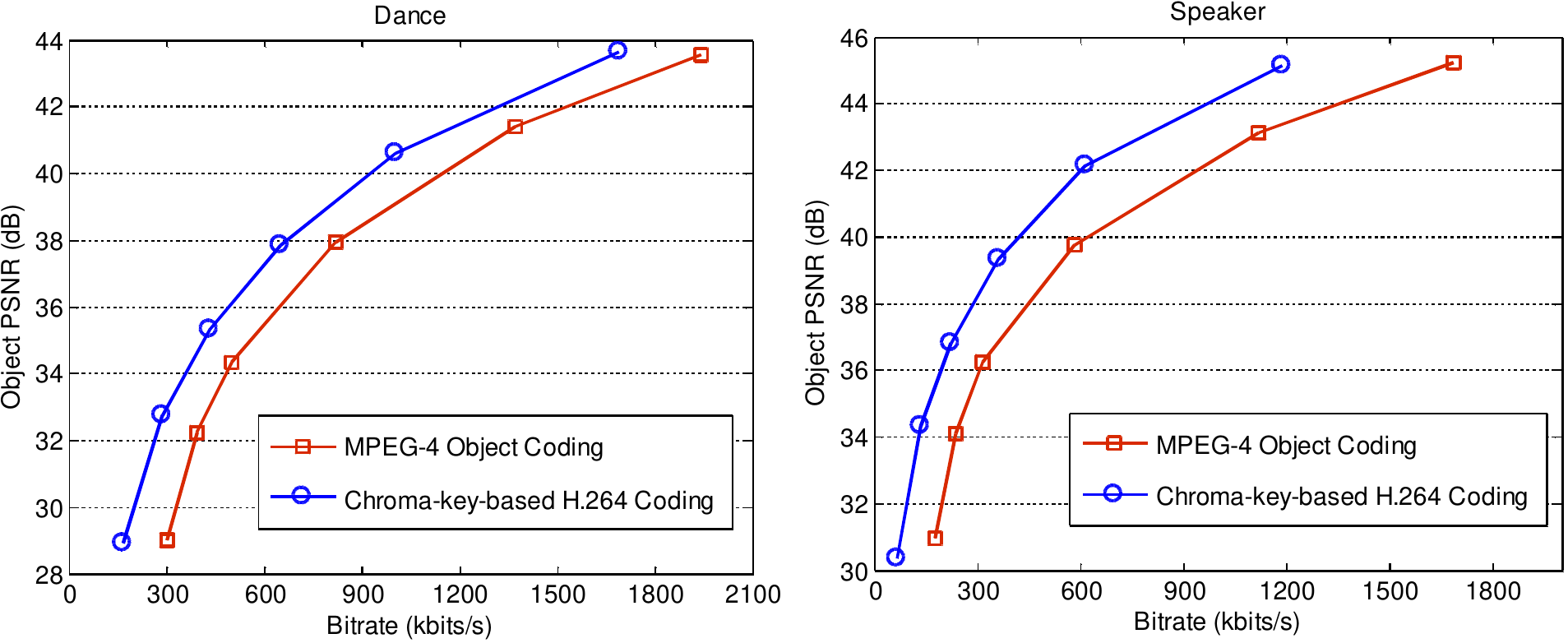}
   \end{center}
   \vspace{-12pt}
   \caption
   {\label{fig:coding1}
	 Performance comparison of implicit shape and object coding using H.264 and MPEG-4.}
	 \vspace{-16pt}
   \end{figure}

\vspace{2pt}
{\bf Implicit H.264 object coding.} 
Although only MPEG-4 is supporting arbitrarily-shaped object coding, we utilize H.264 here due to its higher coding efficiency. For standard compliance, a chroma-key-based scheme using H.264 is adopted over MPEG-4 object coding to encode the cutout object. Background is replaced by a special color prior to encoding and chroma keying is used to recover the object region at the receiver site. Fig.~\ref{fig:coding1} shows the superior coding performance of chroma-key-based H.264 over MPEG-4 object coding. 

   \begin{figure} [t]
   \begin{center}
   \includegraphics[width = 0.8\columnwidth]{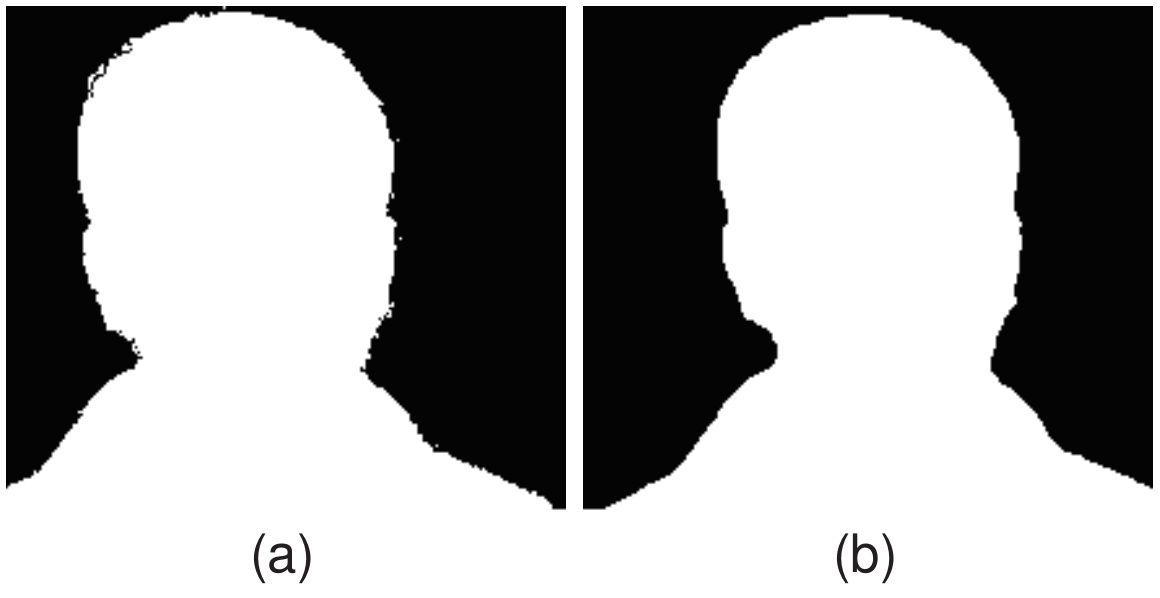}
   \end{center}
   \vspace{-8pt}
   \caption
   {\label{fig:coding2}
	 Chroma keying-based clean shape recovery. a) Using simple thresholding filter. b) Using the proposed nonlinear filter.}
   \end{figure}

\vspace{2pt}
{\bf Effective chroma keying.} Due to quantization artifact, background color can cross the boundary and bleed into the object region. Simple thresholding in the shape recovery will result in many outliers and undesirable boundary (see Fig.~\ref{fig:coding2}). To handle these misclassified pixels, a robust nonlinear filter is applied to the binary mask obtained by the chroma keying. Essentially, a pixel is reassigned as a background pixel if the number of neighboring background pixels exceeds a certain threshold.

{\bf Fast encoding scheme.} 
Due to the advanced coding features, H.264 encoder will consume the precious computing resources, which are also
required by other critical tasks (e.g. video object cutout). Using only a lightweight setup, it is important to
reduce the encoder complexity, leaving more CPU resources for other tasks to realize such a system in real time.

\begin{figure} [t]
   \begin{center}
   \includegraphics[width = 0.85\columnwidth]{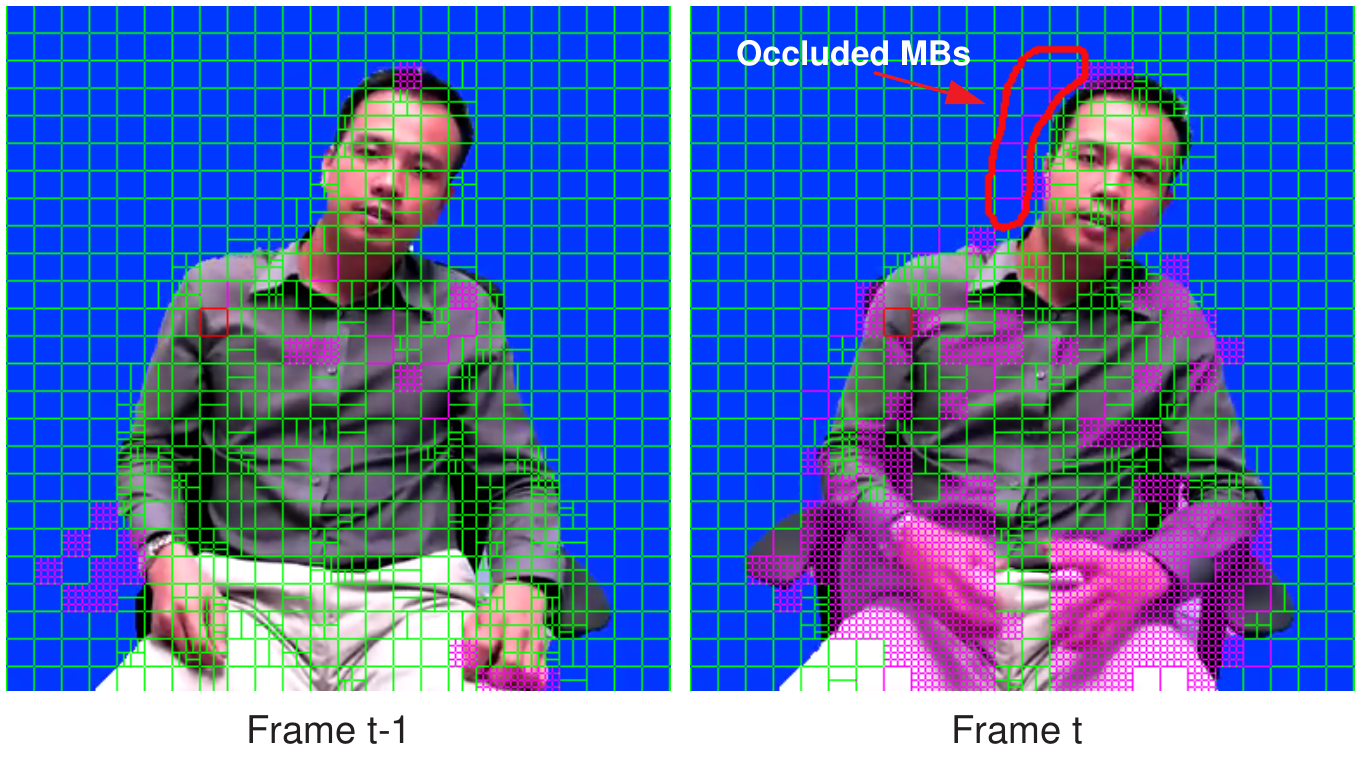}
   \end{center}
   \vspace{-12pt}
    \caption
   {\label{fig:sample_frame}
   Sample frames of a real-life video conferencing sequence with chroma-key background.}
    \vspace{-14pt}
   \end{figure}

Due to the variable block size coding feature, the mode decision process in the standard H.264/AVC encoder is known as one of the most computationally intensive components. To reduce the complexity, many methods have been proposed to selectively examine only a number of highly probable modes \cite{csvt11,csvt09,csvt08}. Unnecessary modes are commonly skipped based on some criteria drawn from a particular observation and behavior in the common test sequences (e.g. Foreman, News). However, lighting condition and unique nature of contents can make real-life video conferencing sequences to exhibit different characteristics; thus, it may be ineffective when applying these fast algorithms. For example, unstable lighting condition and shadow usually vary the luminance channel significantly between successive frames and make it less efficient to exploit the spatial/temporal mode information coherence like in \cite{csvt09,csvt08}. Furthermore, chat videos often contain sudden and complex motion (e.g. hand gesture, face expression), which sometimes exhibits a high degree of motion activity due to a close camera distance. We observed that motion and lighting-affected regions due to shadow in chat videos are generally intra coded and appear in a cluster. Fig.~\ref{fig:sample_frame} shows the sample frames of a chat video, where intra-coded macroblocks (MBs) decided by the full rate-distortion optimization (RDO) process are highlighted in red. The details of the proposed fast object coding algorithm will be described shortly in the next section, where we take into account the characteristics of real-life video conferencing sequences to significantly reduce the encoder complexity while retaining acceptable visual quality. 

\vspace{-8pt}
\subsection{Spatialized Audio Capture and Localization}
One of the key elements in realizing multiparty tele-immersive systems is the ability to spatialize sound and detect speakers, especially in a scenario of multiple participants at a site. It will enable a radically new communication experience with a better intelligibility and an easier comprehension in such scenarios. To achieve this goal, unlike the existing systems requiring a large spatially separated microphone array, we newly build a significantly compact microphone array~\cite{SZhao-ICIEA}, where four collocated miniature microphones are constructed in the approximation of the acoustic vector sensors (AVS) (Fig.~\ref{fig:fig1}). This AVS array consists of three orthogonally mounted acoustic particle velocity gradient microphones X, Y, and Z and one omni-directional acoustic pressure microphone O, which is referred to as the XYZO array. The use of gradient microphones over traditional pure pressure microphones is to exploit more available acoustic information including the amplitude as well as the time difference compared to only the time difference as used in pure pressure microphone arrays. As a result, the XYZO array offers better performance in a much smaller size. In this system, we for the first time deploy and evaluate the XYZO array for the 3D capture and sound source localization (SSL) problems. Our 3D audio capture and reproduction utilizes beam forming techniques for filtering each beam through the corresponding head-related transfer function (HRTF) to emulate human sound localization based on the filtering effects of the human ear. Meanwhile, 3D SSL is based on 3D search on the formulated spatial spectrum of the complete frequency range to estimate direction of arrival (DOA). In group teleconferencing, our system not only supports 3D audio perception but also active speaker detection. With addition of a depth sensor, the visual content of the active speaker can be accurately tracked and segmented by fusing both audio and visual cues. This will enable compelling and effective communication experience by immersively rendering the active speaker with the shared contents (see Fig.~\ref{fig:mp}).

\vspace{-8pt}
\subsection{Networking and Multiparty Structure}
\begin{figure} [t]
   \begin{center}
   \includegraphics[width = 1\columnwidth]{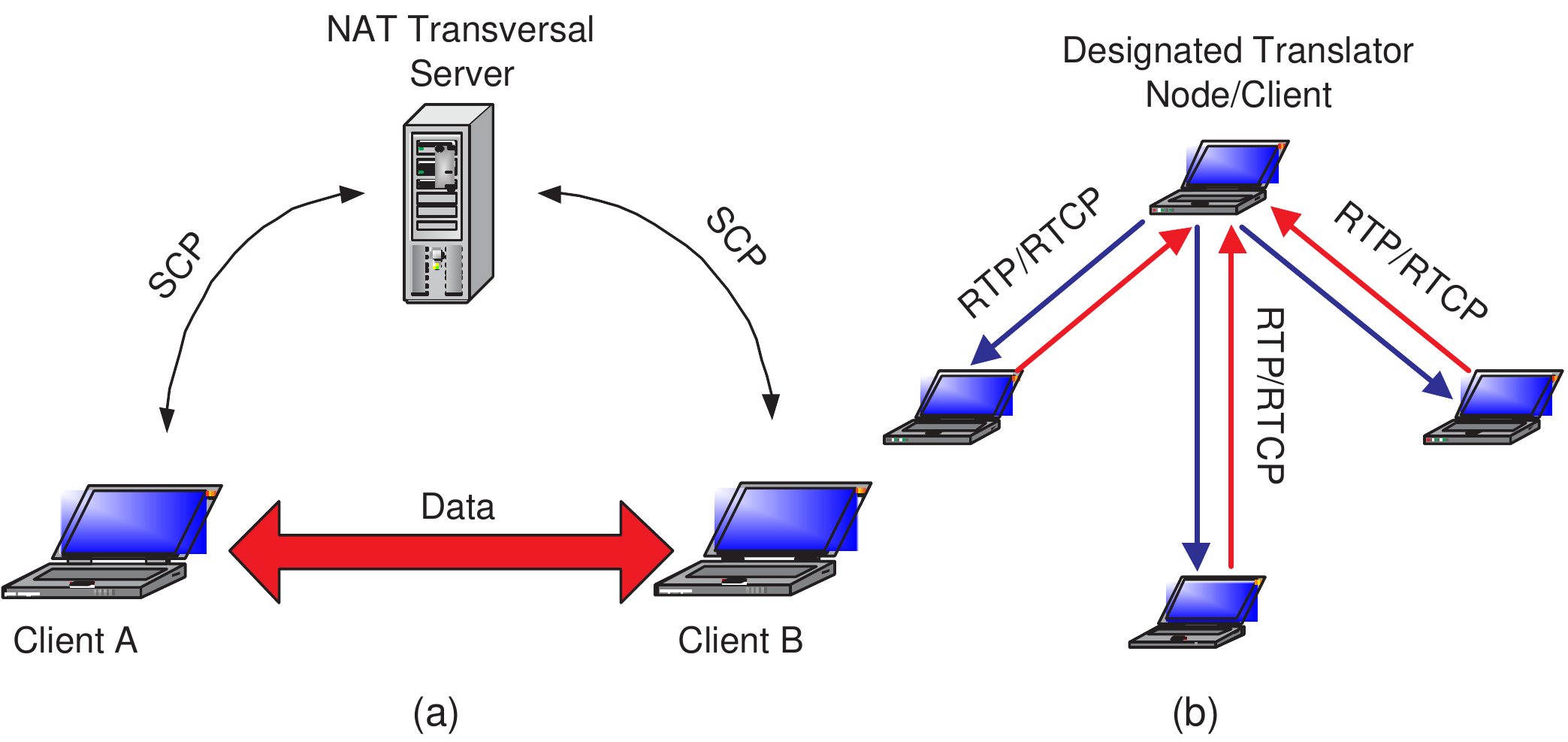}
   \end{center}
   \vspace{-15pt}
   \caption
   {\label{fig:net1} Networking and multiparty architecture. (a) NAT traversal mechanism. (b) Multiparty conferencing architecture in ITEM.}
   \vspace{-15pt}
   \end{figure}

A P2P connection is established between two clients in ITEM for exchanging media data. To traverse network address translation (NAT), a server in public domain with the basic implementation of a UDP hole punching technique is deployed to establish the P2P connection. Actual media data is exchanged in the P2P session without going through the NAT traversal server (see Fig.~\ref{fig:net1}). 

To provide the scalability, our system design supports a mechanism to flexibly specify the transmission structure for media data during a session initialization using SCP. Currently, we support two architectures for data transmission among multiple users: 1) Decentralized ad-hoc structure for a small group meeting, 2) Multicast-based structure for one-to-many connections. In the decentralized ad-hoc structure, we use a node as a designated translator, which establishes P2P connections to other nodes. Each node in a session will only transmit data to the designated translator, which in turn relays the data back to all nodes. The design is simple, inexpensive compared with a centralized solution with a dedicated multipoint control unit (MCU), while it avoids the computing and bandwidth bottleneck with an increased number of concurrent sessions. Compared with a full-mesh connection in~\cite{coliseum,cutechat} the uplink bandwidth at each node is significantly reduced and independent on the number of users, except for the translator node. Meanwhile, the multicast-based  structure is used to support a large number of passive users (e.g. in e-learning), where overlay multicast techniques are employed, if necessary. 
The current design makes it easy to enhance and extend the networking capabilities in future.

\section{Fast Object-Based Video Coding}

To speed up the encoding process, computational power is mostly reserved for the object region. Extensive mode decision process for the background region is eliminated with the early skip mode detection.
In our work, we adopt a coarse-to-fine scheme to determine the best mode, in which some good candidates are identified at a coarser level using a low-complexity criteria and refined later using more optimized criteria. Sharing some similarity with \cite{csvt11}, we use motion cost to determine potential candidates and refine through a RDO process. Specifically, the optimal motion vectors (MVs) of each inter mode are determined by minimizing a motion cost function \begin{equation}
J_{mv} = SATD + \lambda_{mv} R_{mv}	
\end{equation}
where $SATD$ denotes the sum of absolute Hadamard transformed residue, $\lambda_{mv}$ is a Lagrange multiplier, and $R_{mv}$ is an estimated MV cost. However, we evaluate $J_{mv}$ using only $1/2$-pel motion vectors (MVs) and perform $1/4$-pel MV search for only the selected good candidates in the refinement stage to reduce the motion estimation (ME) time. Furthermore, taking into account real-life video conferencing characteristics, we derive more accurate criteria to eliminate unnecessary modes and show a better performance than the existing algorithms.

\vspace{-4pt}
\subsection{Early Skip Mode Detection} The basic principle of identifying a possible skip mode is to examine the cost function. If the cost function is sufficiently small (e.g. below some threshold $Th_\text{SKIP}$), the current MB will be likely encoded as a skip mode. Here, we use the motion cost of a skip mode, $J_{mv}$(SKIP), and adopt the exponential shape function of QP for the threshold as proposed in~\cite{csvt09}. As motion cost is used instead of RD cost, we empirically optimized the function parameters and obtained the threshold as  $\displaystyle Th_\text{SKIP}=14e^{0.1384\times QP}$. Note that this thresholding technique also enables a quick bypass for blue background MBs. However, we also observed that most of blue MBs, which are occluded in previous frames, are intra coded (see Fig.~\ref{fig:sample_frame}). Thus, we propose to early identify and assign only I16MB mode on these MBs.

\vspace{-4pt}
\subsection{Selective Intra/Inter Decision} Existing algorithms tend to assume intra modes are rarely selected as the best mode in inter frames and sometimes incorrectly discard them.  While only negligible quality loss is observed in the standard test sequences, significant rate increment can occur in real-life videos as mentioned above. Here, we propose to evaluate the degree of temporal/spatial correlation of the current MB to early terminate intra/inter mode checking. The spatial correlation degree is measured by a fast intra cost estimation of I4MB and I16MB. We use a low-complexity cost function to determine the optimal mode for each $4\times4$ block, $m_k$, in I4MB as
\begin{equation}
J_\text{I4MB}(m_k) = SATD + \lambda_{md}.4P
\end{equation}
where $P$ is equal to 0 for the most probable intra mode, which is obtained based on the prediction modes of neighboring $4\times4$ blocks, and 1 for the other intra modes. The intra prediction cost for I4MB will be computed as $\displaystyle J_\text{I4MB} = \sum_{k=1}^{16} J_\text{I4MB}(m_k)$. To reduce the complexity, a fast intra mode prediction in our previous work is adopted~\cite{vanguyen}. It is based on the dominant edge 
information that can be fast estimated from the transform coefficients as
\vspace{-0.13in}
\begin{equation}
\vspace{-0.04in}
\theta = \tan^{-1}\left(\sum^3_{j=1}F_{0j} / \sum^3_{i=1}F_{i0}\right) \label{eq5}
\end{equation}
\noindent where $\theta$ is the angle of the dominant edge and $F_{ij}$'s are the transform coefficients of a $4 \times 4$ block. Together with the DC mode, we propose to select additional two out of nine intra-prediction modes, which have the closest orientations to the edge angle $\theta$. 

Meanwhile, the temporal correlation will be measured by the motion cost of a $16\times16$ MB. Let us define 
$J_\text{intra} = \min \{J_\text{I4MB},J_\text{I16MB}\}$. The intra mode candidates can be eliminated  if  $J_{mv}(16\times16) < J_\text{intra}$. As aforementioned, intra MBs in chat videos tend to appear in groups. We can exploit this property to skip examining inter modes. Basically, we consider neighboring MBs (i.e. left, upper, upper-left, and upper-right). If there are at least three intra-coded MBs, it is likely that the current MB will be intra coded if  $J_\text{intra} < J_{mv}(16\times16)$, and we can eliminate all inter modes.

\vspace{-6pt}
\subsection{Candidate Mode Selection} As observed in~\cite{csvt11}, 95$\%$ best RDO mode lies in the subset of three candidate modes with the smallest motion and intra costs. 
We performed extensive simulations on real chat videos and observed that using the motion cost with $1/2$-pel MVs achieved a similar performance with that of 1/4-pel MVs while reducing the ME time. Hence, we shall use $1/2$-pel motion cost to identify potential candidates and eliminate some unlikely inter-mode partitions. 

To reduce ME time, it is critical to early decide if the current MB needs to use $P8\times8$ sub-partitions. We observed that using the criteria $J_{mv}(16\times16) < J_{mv}(8\times8)$ to eliminate $P8\times8$ can provide good performance at high QPs, while causing quality loss at small QPs. This is because at small QPs motion cost tends to overestimate the actual RD cost of small partitions and does not provide effective discrimination. We propose to incorporate a scaling parameter $\beta$ as a function of QP to correct the estimation and use $J_{mv}(16\times16) < \beta J_{mv}(8\times8)$ criteria instead. $\beta$---a monotonically increasing function of QP is empirically obtained through training test sequences such that the probability of not selecting $P8\times8$ as the best mode is 0.85 subject to above criteria.

The overall flow of the proposed method is as follow:
 
\textbf{Step 1:} If  $J_{mv}$(SKIP)$<Th_\text{SKIP}$, select SKIP as the best mode and goto next MB.

\textbf{Step 2:} If $J_\text{intra} > J_{mv}(16\times16)$, eliminate intra modes, otherwise if the number of intra MBs in the neighborhood $\geq$ 3, eliminate all inter modes and goto Step 5.

\textbf{Step 3:} If $J_{mv}(16\times16) < \beta J_{mv}(8\times8)$, eliminate $P8\times8$ and goto Step 5.

\textbf{Step 4:} Determine the partition for each $8\times8$ block in $P8\times8$ mode. If $J_{mv}(8\times8) < \min\{J_{mv}(8\times4), J_{mv}(4\times8)\}$, skip the ME of $4\times4$ partition.

\textbf{Step 5:} Select up to three candidates with the smallest motion cost or intra prediction cost. Perform 1/4-pel refinement for the selected inter-mode candidates and select the best mode through the RDO process.

\section{3D Sound Capture and Localization}
The ITEM system uses our developed low-cost miniature microphone array for 3D binaural audio capture and reproduction. The 3D SSL together with the user tracking using the depth sensor allow automatically detecting and tracking the active speaker for a better communication experience.

\vspace{-5pt}
\subsection{3D Sound Localization}
   \begin{figure} [pt]
   \vspace{8pt}
   \begin{center}
   \includegraphics[width = \columnwidth]{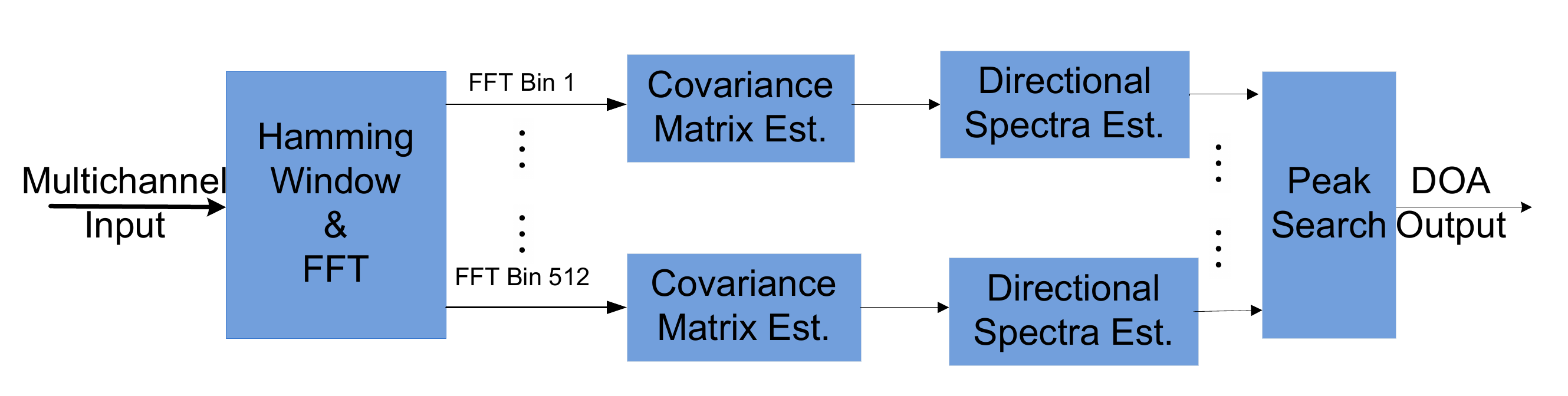}
   \end{center}
   \vspace{-15pt}
    \caption
   {\label{fig:DOA} Workflow of the DOA estimation.}
  \vspace{-13pt}
   \end{figure}
To utilize the XYZO array for the 3D SSL, an offline calibration process is performed once to obtain the steering vectors of the array, which is the impulse response of the array in the 3D (vertical $180^\circ$ $\times$ horizontal $360^\circ$) space. 
The proposed online 3D sound localization module consists of the sound capturing interface, the target sound detection and more importantly the direction of arrival (DOA) estimation. The DOA estimation is implemented as a 3D search on the formulated spatial spectrum of the complete frequency range. 

The overall workflow of the DOA estimation is shown in Fig.~\ref{fig:DOA}. The multichannel input received at the XYZO array is processed by the DOA estimation as follows. The multichannel input is first segmented by a rectangular window, which has a size of 15360 signal samples. Each segment of signal data is termed as a frame. There is one frame for each microphone in the XYZO array, which is equally split into 30 blocks of 512 samples. A Hamming window and a fast Fourier transform (FFT) are applied to each block of 512 samples for converting the data into the frequency domain. For each frequency bin, the covariance matrix is average estimated based on the 30 blocks of data. To estimate the narrowband spectrum for each frequency bin, the inverse of the covariance matrix is calculated and multiplied by the steering vector. The obtained narrowband spectrums are summed across all the frequency bins to give a combined spectrum. The DOA output representing by the horizontal and vertical angles of the sound source is then found by searching the peak along the combined spectrum, which corresponds to the 3D spatial space $ \Phi \in [0^\circ,180^\circ]\times[0^\circ,360^\circ]$. Mathematically, the DOA estimation is represented as 
\begin{equation}
\hat{\theta}_s = \arg\max_{\theta \in \Phi} \sum_{w} f\left ( \hat{R}_X(w,\theta) \right)	
\vspace{-2pt}
\end{equation}
where $\hat{\theta}_s$ denotes the estimated DOA of the sound source, $f (\cdot) $ is a function of a narrowband directional spatial spectra. In this work, we successfully compute this function by using the frequency-domain minimum variance beamformer~\cite{jones}, which has been shown to be robust to background noise. $\hat{R}_X(w,\theta)$ is the covariance matrix of the input vector and it is estimated by the sample-averaging approach.

Generally, the DOA estimation complexity is intensive and increases with a finer search resolution. Currently, without source code optimization, we can achieve a real-time CPU implementation of the 3D SSL with the resolution of $5^\circ$, which allows accurate identification of the active speaker in our field trials as shown later. It should also be noted that with the highly parallel computational structure, a real-time 3D SSL with a finer resolution can be obtained with only an additional GPU that may benefit in a complex meeting setup with a large number of participants. To justify, we realize a GPU implementation of 3D SSL using CUDA programming by parallelizing each 3D search direction and each time-frequency bin using independently different thread blocks. We
implemented both a single-thread and a multi-thread CPU version using
OpenMP~\cite{openmp}. The execution was measured on a four-core Intel Xeon CPU running at 2.7 GHz with the GTX480 for GPU
experiment. For run-time simulation, the GPU implementation achieved 501X and 130X speedup compared to a single-thread and multi-thread CPU implementation with the resolution of $1^\circ$, as shown in Table~\ref{tab:speed}, respectively.  
\begin{table} [t]
\centering
\small
\caption{3D SSL runtime comparison in msec.}
\label{tab:speed}
\begin{tabular}{|c|c|c|c|c|} \hline
  \multicolumn{2}{|c|}{CPU}  & GPU & \multicolumn{2}{c|}{Speedup} \\
\hline
  Single & Multi-thread & & Single & Multi-thread \\ \hline 
 7367 & 1919 & 14.71 & 501X & 130X \\ \hline
\end{tabular}
\vspace{-14pt}
\end{table}
\vspace{-5pt}
\subsection{3D Capture and Reproduction}
The key contribution of the proposed 3D audio capture and reproduction approach is to utilize a compact design microphone array, which is different from the existing approaches. Our approach requires a one-time offline process to determine the optimal time-invariant filter coefficients. These coefficients will be used to convert the B-format signal captured from the XYZO microphone array into the binaural signal for 3D sound playback. In the proposed system, the optimal coefficients are estimated in the frequency domain using the minimum mean squared error (MMSE) approach.   

In concise, the signals received at the left and right ears are the convolutions of the head related impulse response (HRIR) and the incident plane waves. In  the frequency domain, these signals can be modeled as a linear combination of the head related transfer functions (HRTFs) and the incident plane waves, in which the HRTFs are mapped to the incident directions of the plane waves. The number of directions is chosen according to the number of angles considered in the spatial space. The HRTF used for each direction is independent of the number of sound sources. At each frequency bin in the frequency domain, the signal received at each ear, $Y(w)$, can be approximated using a gain vector on an array output vector 
\begin{equation}
\hat{Y}(w) = \textbf{g}^H(w)X(w)
\vspace{-0.03in}
\end{equation}
where $\textbf{g}^H(w)$ is the gain vector and $X(w)$ is the signal received from the XYZO microphone array in the frequency domain. The optimal gain vector can be obtained by solving the MMSE problem as
\begin{equation}
\hat{\textbf{g}}(w) = \arg \min_{\textbf{g}^H(w)} E\left [ \left| Y(w) - \textbf{g}^H(w)X(w) \right|^2 \right]
\end{equation}
It should be noted that the gain vectors are estimated separately for the left and right ears.
In the current implementation, the frequency-domain signals are obtained using a 512-point FFT. As a result, we have 512 gain vectors for each ear signal; each vector contains four elements corresponding to four-channel outputs of the XYZO array. These optimal gain vectors are computed in an offline manner for the real-time system. The optimal time-invariant filter coefficients for each channel are obtained through inverse FFT. 

In a real-time system, the online process consists of signal acquisition, signal segmentation and mapping using the optimal filters, and 3D binaural playback. The sound inputs received at the four microphones are first segmented separately by a rectangular window, which has a size of 512 samples. Each segment of data samples is shifted to a frame with a size of 1024 samples. The convolutions of the filtering are performed in the frequency domain using a 1535-point FFT with zeros padding method. The results of convolutions from each channel are summed to give a combined output for the binaural playback. Since the complexity of the online process mostly involves convolutions, the 3D audio reproduction can perform comfortably in real time on a commodity desktop/laptop.

\vspace{-5pt}
\subsection{Active Speaker Detection}
\begin{figure} [pt]
   \begin{center}
   \includegraphics[width = \columnwidth]{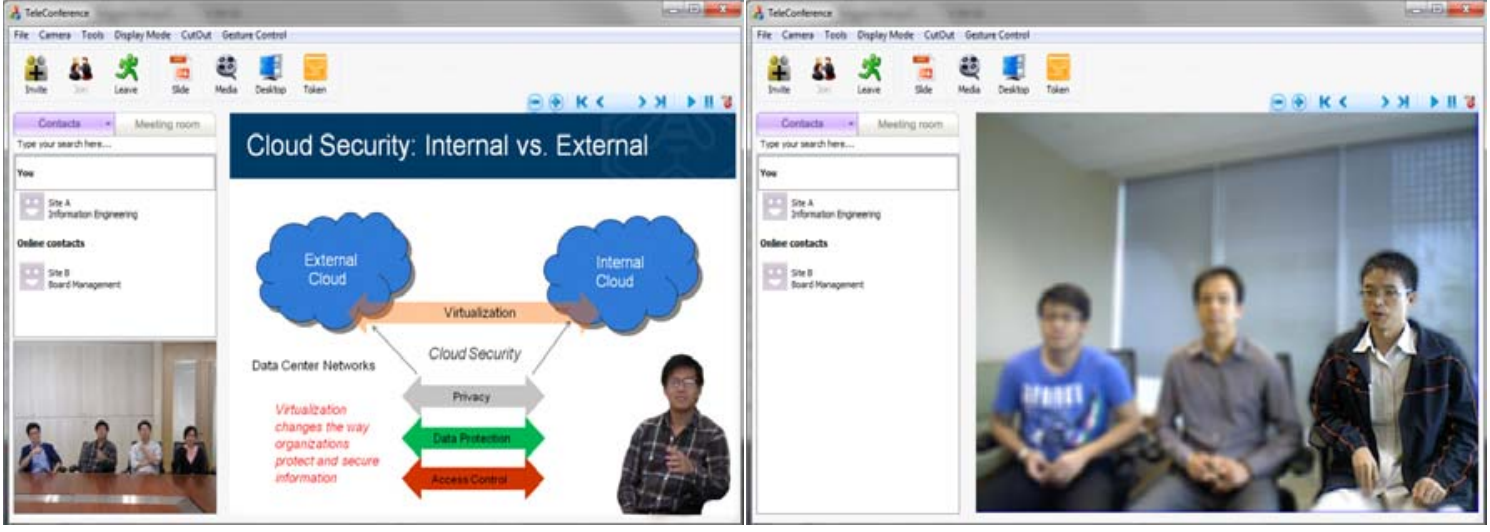}
   \end{center}
     \vspace{-12pt}
    \caption
   {\label{fig:mp}  Active speaker detection with intuitive visual presentation. Left: Presentation mode with the immersive composition of the current speaker into the shared slides. Right: Discussion mode with the focusing effect on the speaker.}
   \vspace{-16pt}
   \end{figure}

The output of the 3D SSL will be used together with the user tracking information obtained through a depth sensor to identify an active speaker. We utilize the OpenNI SDK for depth sensor to track multiple participants in the same physical room at one site and obtain their corresponding visual contents in real time. Specifically, each participant is associated with a user ID, which can be used to extract useful information (e.g. user's joints' 3D positions relative to the camera coordinate, 2D user's image coordinates and user's color segmentation). Thus, the video region of the current speaker can be extracted through his user ID induced by mapping the detected sound source angles from the 3D SSL to the depth sensor's user tracking 3D positions. By fusing cues from different sensors, it also prevents a wrong 3D SSL detection result, if any, when it loses track due to the room noise. Meanwhile, the knowledge of a sound source also helps to retrieve a lost tracking when a participant happens to move out of the scene and return. For more compelling user experience, the active speaker is segmented and immersively composited for various intuitive presentations (see Fig.~\ref{fig:mp} and Section~\ref{sec:sysPerformance}). 

   \begin{figure*} [t]
   \begin{center}
   \includegraphics[width = 1.0\textwidth]{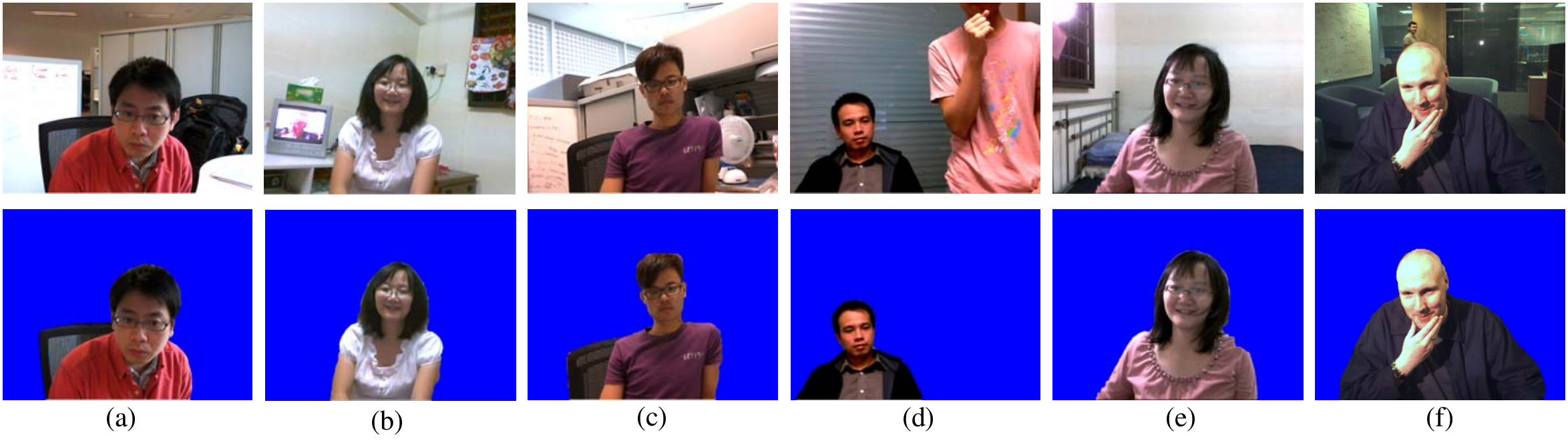}
   \end{center}
   \vspace{-14pt}
   \caption
   {\label{fig:cutdemo}
    Foreground segmentation results for various test cases. (a) A large number of similar black pixels in the background. (b) Picture changes on the TV set and time-varying cast shadow of a ceiling fan. (c) A rotating fan in front of the clutter background. (d) A intruder of bigger size walked toward the camera. (e) A lamp in the up-left corner switched on. (f) A moving person in the background (a video clip from \cite{bi-cvpr06}).}
    \vspace{-12pt}
   \end{figure*}

\section{System performance}

In this section, we evaluated the performance of the key components and the whole system using a commodity setup (a laptop/desktop with a four-core Intel Xeon CPU running at 2.7 GHz), a single webcam, a Kinect depth camera and a compact microphone array. 

\vspace{-5pt}
\subsection{Video Object Cutout}

\begin{table} [b]
\centering
\small
\caption{Execution speed of video object cutout (FPS) using a webcam only with and without multi-threading}
\label{tab:tab1}
\begin{tabular}{|c|c|c|c|c|} \hline
	\multirow{2}{*}{Resolution} & \multicolumn{2}{c|}{w/o multi-threading} & \multicolumn{2}{c|}{w/ multi-threading}  \\ \cline{2-5}
 		& Laptop & Desktop & Laptop & Desktop\\ \hline
    $640\times480$ & 12.3 & 12.1 & 17.1 & 18.4\\ \hline
    $1280\times960$ & 9.6 & 9.3 & 13.9 & 17.5\\ \hline
\end{tabular}
\vspace{-6pt}
\end{table}

\begin{table} [b]
\centering
\small
\caption{Execution speed of video object cutout (FPS) with and without using a depth camera}
\label{tab:tab2}
\begin{tabular}{|c|c| c|c| c|} \hline
	 \multirow{2}{*}{Resolution} & \multicolumn{2}{c|}{webcam only} & \multicolumn{2}{c|}{w/ depth camera}  \\ \cline{2-5}
 	 &	Laptop & Desktop & Laptop & Desktop\\ \hline
   $640\times480$ & 17.1 & 18.4 & 25.4 & 25.9\\ \hline
   \multicolumn{4}{c}{} \\
\end{tabular}
\vspace{-18pt}
\end{table}

Fig.~\ref{fig:cutdemo} shows foreground segmentation results achieved by our video cutout technique under a variety of challenging test conditions. It deals with these cases very well. Table~\ref{tab:tab1} reports the execution speeds of the proposed algorithm tested on a laptop and a desktop PC using only a normal webcam. 
We have measured the speed of using a single CPU core and also that after exploiting the multi-threading (MT) technique to parallelize two selected modules. However, the code has not been fully optimized yet. For the VGA resolution, even our single-core version is able to segment the foreground object accurately at a real-time speed. After MT acceleration, our method can even process $1280\times960$ videos at 14 FPS on the mainstream CPU in one's laptop, while the CPU load is just around 37\%. 
Table~\ref{tab:tab2} reports the execution speeds of video object cutout with and without using the depth camera and MP enabled. Note that we only provided the comparison for VGA resolution, which is currently supported for the color sensor of the Kinect camera. The results show that without fully optimizing the code, our methods can perform comfortably at a real time speed. For all the cases, the CPU load has remained about 20-30\%, leaving much CPU time to process other parallel tasks.

\vspace{-5pt}
\subsection{Object-based Video Coding} 

\begin{table}[t]
\small
\caption{Performance comparison between the proposed method, Lee's method, and Kim's method.}
\begin{small}
\begin{center}
\vspace{-10pt}
\begin{tabular}{@{\ \ } l @{\ \ \ } c @{\ \ \ } c @{\ \ \ }c @{\ \ \ }c @{\ \ \ } c @{\ \ \ } c @{\ \ }}\hline
& \multicolumn{2}{c}{Lee's method} & \multicolumn{2}{c}{Kim's method} & \multicolumn{2}{c}{Our method}\\%
\cline{2-7}
Sequence        & $\Delta$PSNR  & $\Delta$ET  & $\Delta$PSNR  & $\Delta$ET & $\Delta$PSNR & $\Delta$ET \\ 
& (dB) & ($\%$) & (dB) & ($\%$)& (dB) & ($\%$) \\
\hline%
Speaker        									 & -0.321 & -72.1 & -0.258 & -66.3& -0.105 & -71.0 \\ %
Dance   									       & -0.560 & -61.0 & -0.631 & -49.8& -0.231 & -65.1 \\ %
Kelvin (near) \hspace{-0.1in}    & -0.312 & -63.9 & -0.358 & -56.4& -0.191 & -64.1 \\ %
Johan (near) \hspace{-0.1in}     & -0.295 & -65.8 & -0.423 & -58.2& -0.167 & -65.3 \\ %
Kelvin (far) \hspace{-0.1in}     & -0.344 & -62.7 & -0.375 & -53.0& -0.173 & -64.1 \\ %
Johan (far) \hspace{-0.1in}      & -0.326 & -64.5 & -0.381 & -53.8& -0.184 & -66.8 \\ %
\hline
Average    										   & -0.360 & -65.0 & -0.404 & -56.3& -0.175 & -66.1 \\ %
\hline
\end{tabular}
\vspace{-12pt}
\end{center}
\end{small}
\label{tab:exp}
\end{table}

We evaluated the proposed fast object coding method using real-life test sequences in a typical video conferencing scenario. The video object test sequences are segmented from a live webcam, where some screen shots are shown in the second row of Fig.~\ref{fig:cutdemo}. For comparison, we also implemented the methods proposed by Lee~\cite{csvt11} and Kim~\cite{csvt08}.
Table~\ref{tab:exp} shows the experimental results obtained by different fast coding methods. The proposed method reduced the mode decision time by $66.1\%$ while incurring only marginal PSNR loss (0.17-dB). The proposed method also achieved a better performance than the exisiting methods. Though examining mostly more than one mode in the RDO refinement stage compared with Lee's method, using 1/2-pel motion cost and exploiting the cluster characteristics of intra-coded regions enabled the proposed method to achieve a slightly higher complexity reduction. 
As can be seen, designed by using the standard test sequences, the existing methods were not effective when applied for the real-life test sequences. In particular, although more computationally intensive, Kim's method could not provide a better compression performance compared with the proposed method due to the inefficient mode elimination scheme using the temporal mode correlation, which was low due to the sudden and complex motion such as hand gesture at a close camera distance and unstable lighting condition and shadow. Furthermore, the quality degradation in the proposed method is uniformly distributed over a wide bitrate range while Lee and Kim's methods generally resulted in a higher quality loss at a high bitrate range. 

\begin{figure} [t]
   \begin{center}
   \includegraphics[width = \columnwidth]{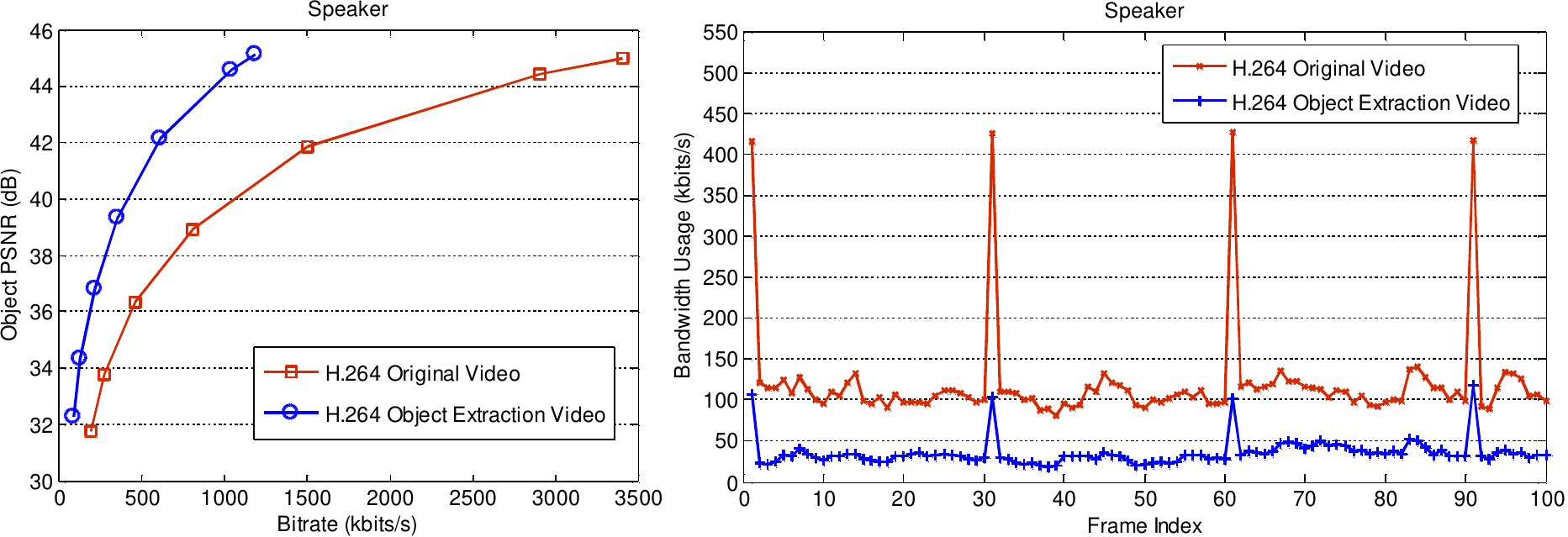}
   \end{center}
   \vspace{-12pt}
   \caption
   {\label{fig:coding3}
	 Bandwidth savings given by object cutout coding compared with the original video content.}
	 \vspace{-14pt}
   \end{figure}

In addition, by throwing out the undesired background and non-interesting objects, it allows both to reduce the required bitrate and to speed up the encoding process. Fig.~\ref{fig:coding3} shows the rate-distortion curves obtained by encoding the original video content and object cutout, respectively. For the same visual quality, about {\it three} times bandwidth savings can be achieved at the medium bitrate range; moreover, smaller sizes of I frames (spikes in the right figure) will ease the network congestion and delay in transmitting video contents. Table~\ref{tab:coding} reports the coding speeds for different video quality with and without object cutout and with and without using the fast mode decision. As can be seen, the encoding process is significantly accelerated, hence leaving more CPU resources for other tasks running concurrently. Compared to \cite{cutechat}, we achieved about $41\%$ and $23\%$ faster execution speeds with our depth-based video object cutout and fast mode decision in object coding. 
\begin{table}[t]
\centering
\small
\caption{Video coding performance comparison}
\vspace{-2pt}
\label{tab:coding}
\begin{tabular}{|c|c|c|c|c|} \hline
	\multirow{3}{*}{Resolution} & Bitrate & \multicolumn{3}{c|}{Encoding frame rate (fps)}  \\ \cline{3-5}
 		&	(Kbits/s) & \multicolumn{2}{c|}{Object cutout} & Original  \\ \cline{3-4}
 		&	 & w/o FMD & w/ FMD &frames  \\ \hline
  \multirow{2}{*}{$640\times480$} & 1,200 & 34.3 & 42.5 & 12.4 \\ \cline{2-5}
                                    & 600 & 41.2 & 50.7 & 16.5 \\ \hline
  \multirow{2}{*}{$1280\times960$} & 2,400 & 16.2 & 19.2 & 6.7 \\ \cline{2-5}
                                   & 1,200 & 18.4 & 22.5 & 8.3 \\ \hline    
\end{tabular}
\end{table}

\vspace{-8pt}
\subsection{Spatialized Audio and Speakers Detection using Multi-sensor Fusion}
\label{sec:sysPerformance}
In case of group teleconferencing with multiple participants at a site, we observed the 3D SSL module could detect the angle of a speaker with the accuracy of $6^\circ$ in real time. Together with the user tracking information obtained through a depth sensor, this always led to accurate active speaker identification and his/her correct video content extraction. It is worth mentioning that the speaker video distractingly switched too often whenever there was a noise sound source within a short period. To eliminate this undesired effect, we used a threshold to verify the 3D SSL new detection output. The system only switches to the new speaker video when a consistent detection output is presented within a certain period.

   \begin{figure} [t]
   \begin{center}
   \includegraphics[width = 0.9\columnwidth]{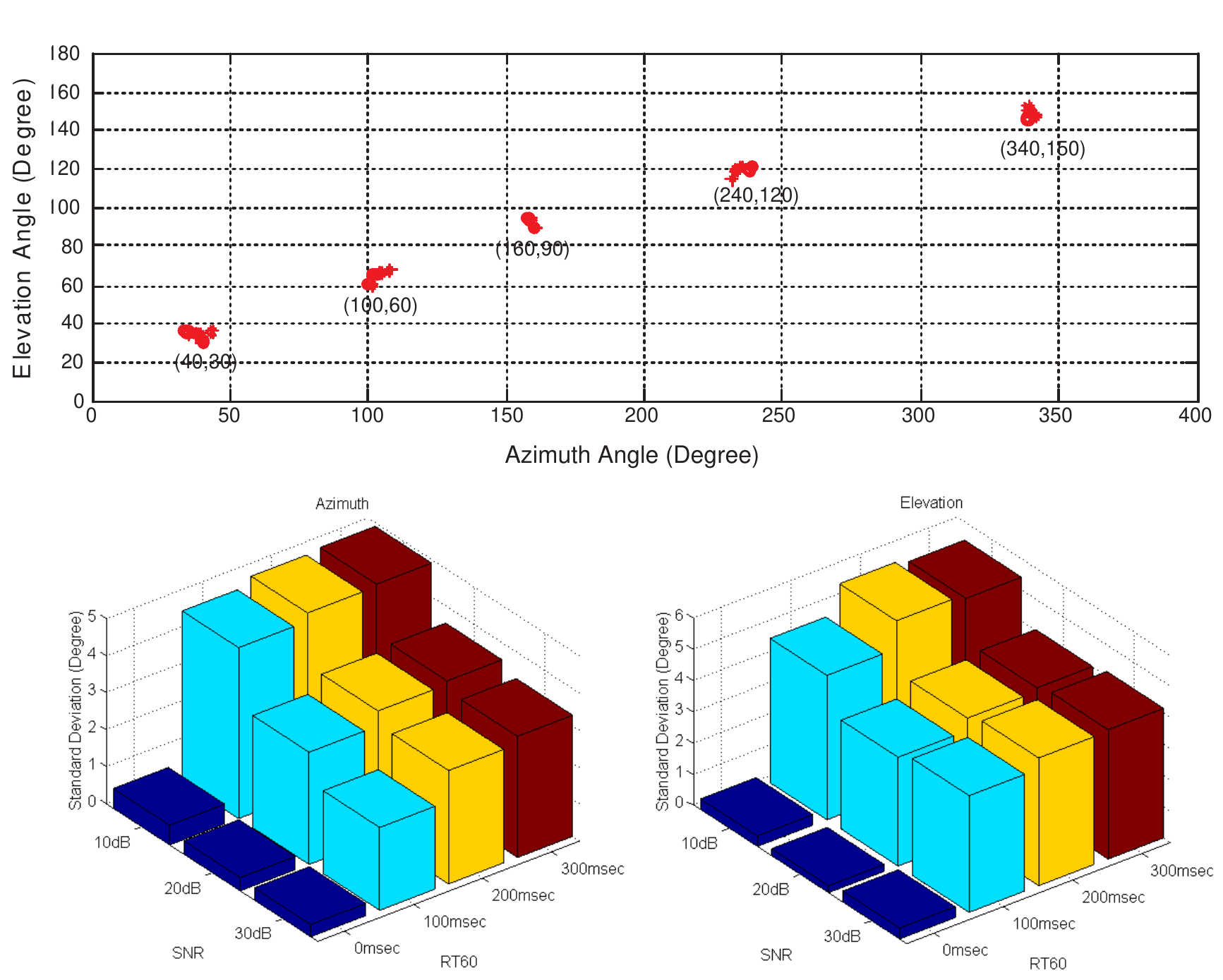}
   \end{center}
   \vspace{-14pt}
   \caption
   {\label{fig:doaAnalysis}
	 The DOA estimation results. Top: Mean values. The values in the brackets are the tested sound source angles. Bottom: Mean standard deviations.}
	 \vspace{-14pt}
   \end{figure}

\textbf{3D Sound Localization.} We evaluated the performance of the proposed 3D SSL with various noise and reverberation levels. Different reverberations were obtained by convolving varying clean speech signals with the measured impulse response of different lengths corresponding to different reverberation times. The tested reverberation lengths were 0msec, 100msec, 200msec, and 300msec. To get different noise levels, an additive white Gaussian noise was added to the resulted signals. The tested noise levels measured by SNR are 10dB, 20dB, and 30dB. Six testing directions of sound sources were randomly selected from the 3D space for all the conditions. For 20 testing realizations, the mean localization results over the total realizations and the mean standard deviations over the total tested directions are illustrated in Fig.~\ref{fig:doaAnalysis}. It is observed that the proposed approach provided the estimated mean azimuth and elevation angles within a small range of the target positions for all the tested conditions. In addition, the mean standard deviations are less than $1^\circ$ for the reverberation time RT60 = 0msec and less than $6^\circ$ when
the SNR decreases to 10dB and the reverberation time RT60 increases to 300msec. The results also show that the proposed DOA method offers considerable robustness to additive noise, but it tends to degrade the performance in reverberant conditions. The overall
results indicate that the localization accuracy of the XYZO array is within 6◦ error in a relatively noisy and reverberant meeting room.

\textbf{3D Binaural Audio Reproduction.} To evaluate the 3D audio reproduction accuracy of the proposed system, psychological subjective tests were carried out. There
were five subjects involved in the evaluation. The assessment criterion is the ability to localize audio sources. In the experiment, the tested audio sources were recorded as
follows: for $0^\circ$ elevation angle there were 24 positions recorded for
the azimuth angles varying from $0^\circ$ to $360^\circ$ with $15^\circ$
interval, and for $0^\circ$ azimuth angle there were 24 positions recorded for the elevation angles varying from $-90^\circ$ to $90^\circ$. The azimuth localization and elevation localization were tested separately. During each
test, the subject was required to hear the processed audio signals
and point to the direction of the audio source. A score of accuracy over 24 positions
was obtained by comparing to the original directions of the audio sources. 
Table~\ref{tab:SSLResolutions} shows the average scores over all the subjects for different reproduction resolutions as $1^\circ$, $15^\circ$, $30^\circ$, and $45^\circ$. The results show the effectiveness of the proposed system, in which the average scores over all the subjects were 20 for 24 azimuth positions and 10 for 24 elevation positions. It is observed that the sound localization for the azimuth positions, which is a more important factor in our system, was better than that for the elevation positions. 

\begin{table} [t]
\centering
\small
\caption{Subject SSL scores for various resolutions.}
\vspace{-4pt}
\label{tab:SSLResolutions}
\begin{tabular}{|l| c c c c | c c c c |} \hline
 & \multicolumn{4}{c |}{Azimuth}  & \multicolumn{4}{c| }{Elevation} \\
 \cline{2-9}
 & $\;1^\circ$ & $\;15^\circ$ & $\;30^\circ$ &  $\;45^\circ$ & $\;1^\circ$ & $\;15^\circ$ & $\;30^\circ$ & $\;45^\circ$   \\
  \hline 
Score & 20 & 20 & 12 & 8 & 10 & 9 & 5 & 2\\ \hline
\end{tabular}
\vspace{-15pt}
\end{table}

\textbf{Active speaker detection.} A real-life experiment was conducted to evaluate the performance of the active speaker detection and his/her video extraction. We invited up to six participants to join a group meeting. They were placed in the same room and each took turns to speak. The simulation results show that the 3D SSL module could detect the angle of a speaker with the accuracy of $6^\circ$ in real time. Together with the user tracking information obtained through a depth sensor, this always led to an accurate active speaker identification and his/her correct video content extraction. It is worth mentioning that the speaker video distractingly switched too often whenever there was a noise sound source within a short period. To eliminate this undesired effect, we used a threshold to verify the 3D SSL new detection output. The system only switches to the new speaker video when a consistent detection output is presented within a certain period.

\vspace{-5pt}
\subsection{Overall system performance} To evaluate the overall system performance, we conducted multiparty conferencing over the public Internet using the decentralized ad-hoc structure. With the compressed video bitrate of 500 kbits/s, ITEM can easily support up to six participants within a session. The typical end-to-end lattency is reported in Table~\ref{tab:latency}. We also measured the total CPU usage of about $35\%$-$40\%$ (about $15\%$ for video object cutout, $10\%$ for video coding/decoding and rendering, $10\%$ for other tasks). With an increased number of participants in a session, we observed about $10\%$-$15\%$ increase in CPU workload (for ten connections over the LAN network) that is consumed by the decoding and rendering processes. The results show that our system has better scability performance compared with \cite{coliseum}, where almost $100\%$ CPU is utilized for only about three participants, leading to a significant low frame rate with an increased number of participants. 
   
\begin{table} [t]
\small
\centering \caption{{Analysis of latency (ms)}}
\vspace{-4pt}
\label{tab:latency}
\begin{tabular}{|l|c|} \hline
Video object cutout & 38-54 \\ 
Video object encoding/decoding & 24-38 \\
Network (jitter, relay, etc.) & 28-43 \\
Rendering and display & 12-30 \\ \hline
End-to-end latency & 102-165 \\ \hline
\end{tabular}
\vspace{-14pt}
\end{table}   

\section{Applications and Case Study}
\label{sec:expe}
This section presents the potential utilization of ITEM to enhance the communication experience and effectiveness in various business and consumer solutions through an appropriate configuration and customization to meet the practical requirements of application scenarios. These potential applications share the same unique and key added feature provided by ITEM where all remote parties are put in a same shared environment, stimulating sense of being together.  

\vspace{-8pt}
\subsection{Business group meeting}
\begin{figure} [t]
\begin{center}
\includegraphics[width = \columnwidth]{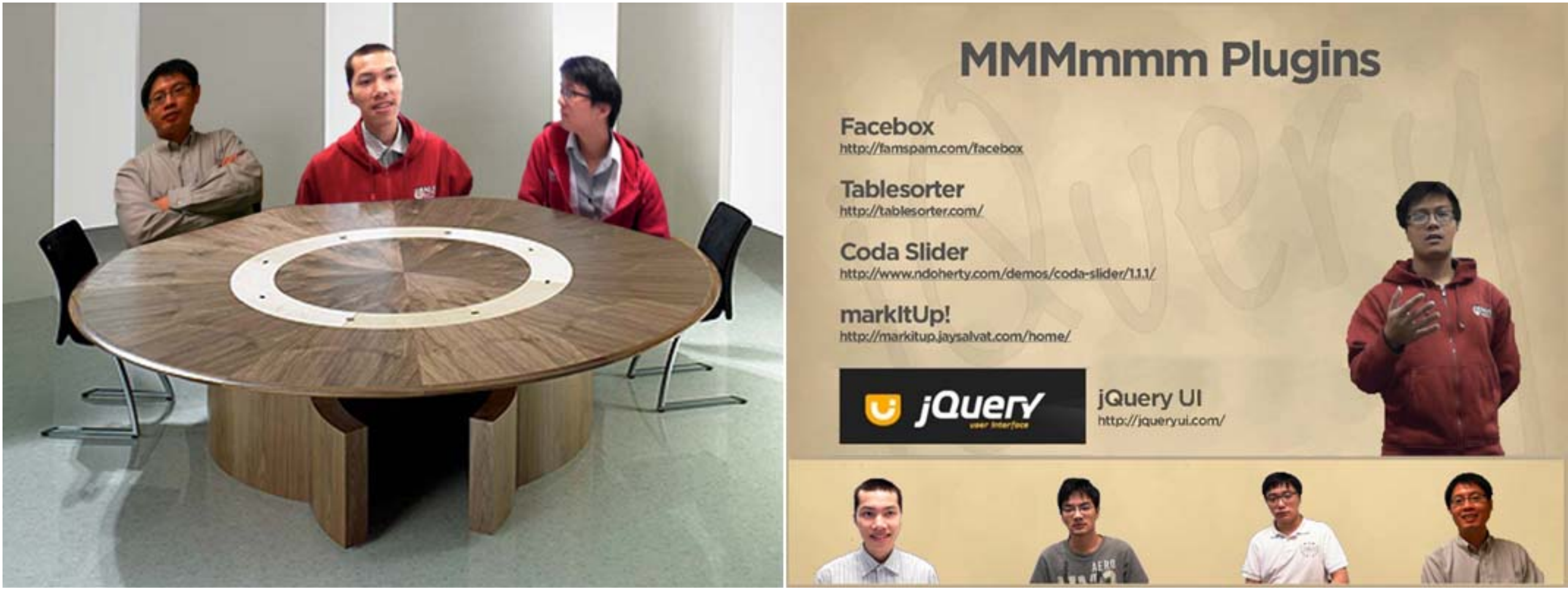}
  \end{center}
  \vspace{-12pt}
    \caption { \label{fig:meeting}Multiparty immersive video communication for an effective business online meeting.}
    \vspace{-16pt}
\end{figure}

Stimulating in-person meeting characteristics (e.g. interaction among participants and collaborative contents with a sense of belonging to the same place) with scalability support is the critical element for effective tele-meeting. We discuss here some choices when configuring and customizing ITEM to realize such required characteristics. For networking structure, we use the decentralized ad-hoc structure to: 1) easily support concurrent meeting scalability, and 2) reduce total bandwidth within a session. Furthermore, ITEM supports a variety of collaborative contents from slides, documents, media-rich contents, and even desktop windows through the meta channel. To support TI functionalities, we currently customize two rendering modes to naturally put participants in the same designed virtual meeting space or shared contents, allowing participants a freedom to navigate around (Fig.~\ref{fig:meeting}). \textit{Presentation mode:} the active speaker is segmented and immersively merged with the shared contents. \textit{Discussion mode:} without any shared contents, the entire remote room is displayed with the active speaker in focus and the blurring effect for other participants. We have deployed our system for the internal trials and collected some initial useful feedbacks. Users like the virtual meeting room design that gives them a strong sense of presence in the same physical space without any distracting, private backgrounds. Although users are not fully satisfied with the current layout in the content sharing mode when having many remote participants, this mode is often preferred due to the need of sharing collaborative contents during a meeting and its effectiveness for conveying the gesture signals to the shared contents. It is observed that when there is a single participant at a location, the user prefers a simple setup of a webcam without the need of using a depth camera for the gesture-based control. Users liked the speaker detection and immersive visual composition features in the presentation mode. They felt the quality of the meeting was more effective with the ease to keep track of both the slide contents and the speaker video at the same time. They commented that the engagement of the active speaker with the shared contents also led to a more effective presentation and discussion. The survey results also show that the 3D audio reproduction feature was more useful in the discussion mode than the presentation mode. In the presentation mode, the main focus was usually on the shared contents and the current speaker. Therefore,
users commented that they did not strongly perceive and pay much attention to the spatialized audio. 

\vspace{-8pt}
\subsection{Distance learning and education}

Fusion of users into the shared contents and virtual spaces in real time offers a variety of uses in distance learning and education. By merging a lecturer into the course materials, the gesture signals of a lecturer and her interactions with the slides are easily perceived as in a physical classroom, which is an important factor for an effective e-learning course. Inviting another lecturer from a different location to join and discuss during a live lecture is made possible by simply adding her as an active ITEM client. In a long-distance training course (e.g. yoga or rehabilitation), a proper rendering mode is designed to effectively put participants in a virtual class room, allowing a learner to easily see and follow the trainer's movement that is hardly feasible or effective with the existing commodity video communication systems. When natural interaction with the course contents or full-body tracking is required, a depth camera is recommended for the active ITEM participants. To support a large number of students (passive ITEM participants) in an e-course, multicast-based networking structure is employed. For the case of yoga or rehabilitation training or a small virtual class room requiring interactive collaborations, the decentralized ad-hoc structure is employed instead.
\vspace{-8pt}
\subsection{Entertainment}

We have deployed ITEM as a light-weight tele-immersive video chat application to bring fun, exciting additions to the video chat experience. The system allows friends and long-distance family members to experience a sense of togetherness by creating a virtual space to let them see, interact, and do something fun together. Being able to separate an user from the background, the application lets users change the background, swap in another, and apply cool, fun video effects such as blurring the background or stylizing the user video (Fig.~\ref{fig:cutout} - bottom-right image). With a similar, simple setup as any video chat application in the consumer space, our application creates endless exciting uses with just one click from sharing vacation photos, covering up a messy room, pretending at some places, or hiding someone else who is watching and listening. We have also allowed users easily creating a fun, immersive video clip and share with friends and relatives on social networks. Creating a photo, video clip with all family members becomes easier than ever regardless of their distant locations.

\begin{figure} [t]
   \begin{center}
   \includegraphics[width = \columnwidth]{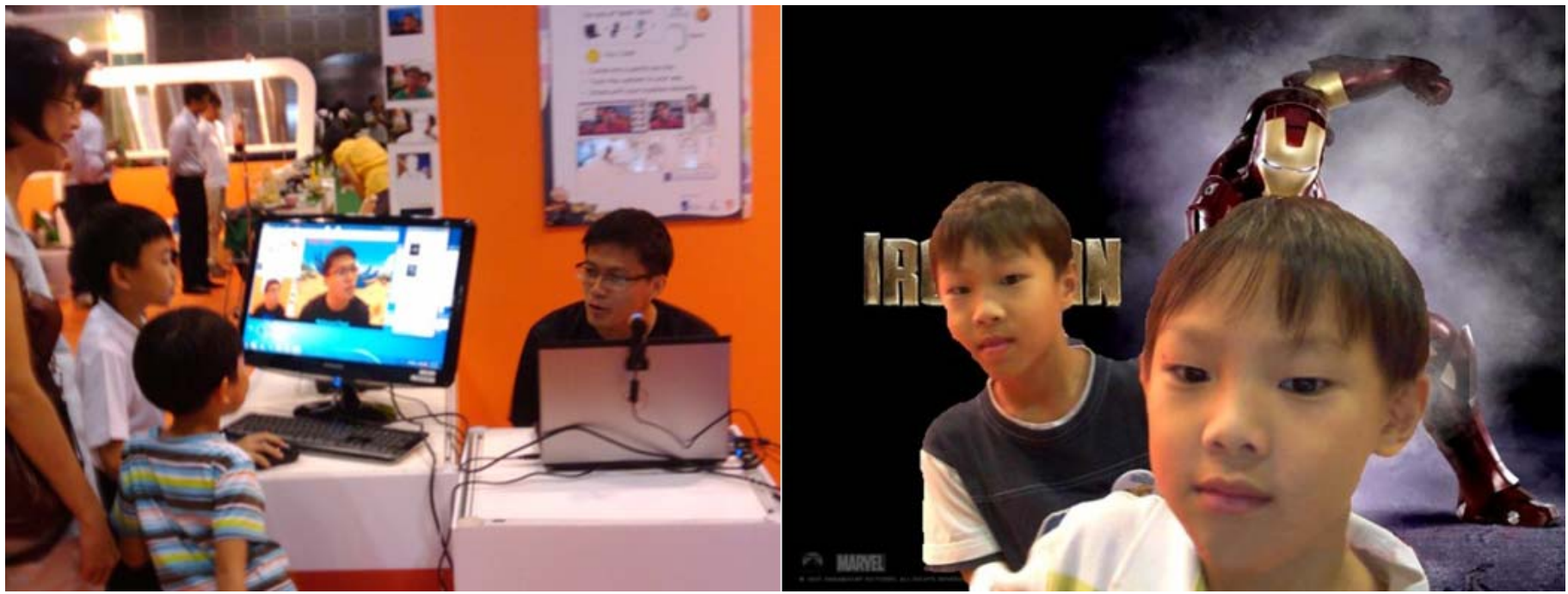}
   \end{center}
   \vspace{-13pt}
   \caption
   {\label{fig:mp1}
    Demo at TechFest for children where the kids were excited and enjoyed themselves with our video chat application.}       
     \vspace{-16pt}
   \end{figure}

We have demonstrated our application at various Technical Festivals (Fig.~\ref{fig:mp1}) and conducted user experience surveys. The results show users like this new feature of instantly sharing something fun, exciting as the background while conducting video chats at the same time. They are impressed by the real-time segmentation of foreground at high quality from live videos, feeling that his/her webcam has been transformed into an intelligent one magically. Users also really enjoy the immersive video chat features, in which they feel more tightly connected with remote friends and selected background contents. Being simple and fun, the application attracts much attention from users (especially children) and gets them involved longer in a long-distance video chat. We find this observation is consistent with a recent HCI study~\cite{ storyvisit-nokia}.


\section{Conclusion}
\label{sec:conc}

In this paper, we have presented the complete design and realization of an immersive telepresence system for entertainment and meetings based on truly commodity capturing, computing, and networking setups. We addressed the challenges in the key components of the proposed ITEM system, but also exploited the synergy between different modules to maximize the overall system performance. We objectively demonstrated how to design and fit all components with the stress of computing resources and different application requirements. The realization of the ITEM system is indeed a proof of feasibility for such a lightweight practical system, in which the performance is verified through field trials. This allowed us an opportunity to conduct an extensive evaluation of user experience both objectively and subjectively. Our future plan is to further improve the current system and related key technology to ultimately make the TI system practical and accessible to massive users. In fact, the constructed TI system could also be served as an open platform for the community to develop and integrate the related novel technologies for the key system components, which include but are not limited to video object cutout, networking, and visual presentation. For this purpose, we plan to release the back-bone system software under a free software license. In addition, we also intend to enlarge and publish the data set of the real world object video sequences, which will be highly useful for the video coding research community. 

\bibliographystyle{IEEE}

\begin{thebibliography}{99}

\bibitem{TIreview}
J. G. Apostolopoulos, P. A. Chou, B. Culbertson, T. Kalker, M. D. Trott, and S. Wee, ``Road to immersive communication,'' {\em in Proc. of the IEEE}, 2012.

\bibitem{cisco} Cisco {T}ele{P}resence. http://cisco.com/telepresence

\bibitem{polycom} Polycom {T}ele{P}resence. http://www.polycom.com

\bibitem{3DAudio}
K.~Inkpen, R. Hegde, M. Czerwinski, and Z. Zhang, ``Exploring spatialized audio and video for distributed conversations,'' {\em in Proc. of ACM Multimedia}, 2010.

\bibitem{mm04}
D. E. Ott and K. Mayer-Patel, ``Coordinated multi-streaming for 3d tele-immersion,'' {\em in Proc. of ACM Multimedia}, 2004.

\bibitem{SSL1}
H.~Wang and P.~Chu, ``Voice source localization for automatic camera pointing system in videoconferencing," \emph{in Proc. of IEEE Conf. Acoustics, Speech and Signal Processing (ICASSP)}, 1997.

\bibitem{SSL2}
R. {Cutler~\em{et~al.}}, ``Distributed meetings: A meeting capture and broadcasting system," \emph{in Proc. of ACM Multimedia}, 2002.

\bibitem{SSL3}
J.~H.~Dibiase, ``A high-accuracy, low-latency technique for talker localization in reverberation environment using microphone arrays," \emph{Ph.D. dissertation, Brown Univ.}, 2001.

\bibitem{Beingthere}
C. Kuster, N. Ranieri, Agustina, H. Zimmer, J.C. Bazin, C. Sun, T. Popa, and M. Gross, ``Towards next generation 3D teleconferencing systems,'' {\em in Proc. of 3DTV-CON}, 2012.

\bibitem{teevee-mm09}
W. Wu, R. Rivas, A. Arefin, S. Shi, R. Sheppard, B. Bui, and K. Nahrstedt, ``Mobile{TI}: A Portable Tele-Immersive System,'' {\em in Proc. of ACM Multimedia}, 2009.

\bibitem{teeve05}
Z. Yang, Y. Cui, B. Yu, J. Liang, K. Nshrstedt, S. H. Jung, and R. Bajscy, ``{TEEVE}: The next generation architecture for tele-immersive environments,'' {\em in Proc. of 7th ISM}, 2005.

\bibitem{item}
V. A. Nguyen, T. D. Vu, H. Yang, J. Lu, and M. N. Do, ``ITEM: Immersive telepresence for entertainment and meetings with commodity setup,'' {\em in Proc. of ACM Multimedia}, 2012.

\bibitem{objcoding}
V. A. Nguyen, J. Lu, and M. N. Do, ``Efficient video compression methods for a
lightweight tele-immersive video chat system,'' {\em in Proc. of IEEE Int. Symposium on Circuits and Systems (ISCAS)}, 2012.

\bibitem{vms02}
C. W. Lin, Y. J. Chang, C. M. Wang, Y. C. Chen, and M. T. Sun, ``A standard-compliant virtual meeting system with active video object tracking,'' {\em EURASIP Journal on Applied Signal Processing}, 2002.

\bibitem{coliseum}
H. {Baker~\em{et~al.}}, ``Understanding performance in {C}oliseum, an immersive videoconferencing system,'' {\em ACM TOMCCAP}, 2005.

\bibitem{cutechat}
J. Lu, V. A. Nguyen, Z. Niu, B. Singh, Z. Luo, and M. N. Do, ``Cute{C}hat: {A} lightweight tele-immersive video chat system,'' {\em in Proc. of ACM Multimedia}, 2011.

\bibitem{bgdcut06}
J. Sun, W. Zhang, X. Tang, and H.-Y. Shum, ``Background cut,'' {\em in Proc. of ECCV}, 2006.

\bibitem{bi-cvpr05}
V. Kolmogorov, A. Criminisi, A. Blake, G. Cross, and C. Rother, ``Bi-layer segmentation of binocular stereo video,'' {\em in Proc. of CVPR}, 2005.

\bibitem{bi-cvpr06}
A. Criminisi, G. Cross, A. Blake, and V. Kolmogorov, ``Bilayer segmentation of live video,'' {\em in Proc. of CVPR}, 2006.

\bibitem{tofcut10}
L. Wang, C. Zhang, R. Yang, and C. Zhang, ``TofCut: towards robust real-time foreground extraction using a time-of-flight camera,'' {\em in Proc. of 3DPVT}, 2010.

\bibitem{csvt11}
J. Y. Lee and H. Park, ``A fast mode decision method based on motion cost and intra prediction cost for H.264/AVC," \emph{IEEE Trans. Circuits and Sys. Video Tech.}, 2011.

\bibitem{csvt09}
H. Zeng, C. Cai, and K. K. Ma, ``Fast mode decision for H.264/AVC based on macroblock motion activity," \emph{IEEE Trans. Circuits and Sys. Video Tech.}, 2009.

\bibitem{csvt08}
B. G. Kim, ``Novel Inter-Mode Decision Algorithm Based on Macroblock (MB) Tracking for the P-Slice in H.264 AVC Video Coding," \emph{IEEE Trans. Circuits and Sys. Video Tech.}, 2008.

\bibitem{vanguyen}
V. A. Nguyen and Y. P. Tan, ``Efficient video transcoding from H.263 to H.264/AVC standard with enhanced rate control," \emph{EURASIP Journal on Advances in Signal Processing}, 2006.

\bibitem{storyvisit-nokia}
H. {Raffle~\em{et~al.}}, ``Hello, is {G}randma there? {L}et's read! {S}tory{V}isit: {F}amily video chat and connected e-books," \emph{in Proc. CHI}, 2011.

\bibitem{SZhao-ICIEA}
S.~Zhao, A.~Ahmed, Y.~Liang, K.~Rupnow, D.~Chen, and D.~L.~Jones, ``A real-time 3D sound localization system with miniature microphone array for virtual reality," \emph{in Proc.  of  IEEE Conf. Industrial Electronics and Applications (ICIEA)}, 2012.

\bibitem{SZhao-DAFx}
S.~Zhao, R.~Rogowski, R.~Johnson, and D.~L.~Jones, ``3D binaural audio capture and reproduction using a miniature microphone array," \emph{in Proc. of Digital Audio Effects (DAFx)}, 2012.

\bibitem{va-acm2013}
V.~A.~Nguyen, S.~Zhao, T.~D.~Vu, D.~L.~Jones, and M.~N.~Do, ``Spatialized audio multiparty tele-conferencing with commodity miniature microphone array," \emph{in Proc. ACM Multimedia}, 2013.

\bibitem{jones}
D.~Jones, ``Beamformer performance with acoustic vector sensors in air," \emph{in Jour. Acoust. Soc. Am.}, 2006.

\bibitem{roundtable}
C. Zhang, D. Florencio, D.E. Ba, and Z. Zhang, ``Maximum likelihood sound source localization and beamforming for directional microphone arrays in distributed meetings," \emph{in Trans. Multimedia}, 2008.

\bibitem{openmp} http://openmp.org/wp/

\end{thebibliography}
\label{sec:bib}

%

\end{document}